\documentclass[prb,aps,preprint,showkeys,showpacs]{revtex4}
\usepackage{amsmath}
\usepackage{amsfonts}
\usepackage{graphicx}
\usepackage{subfig}
\usepackage{booktabs}
\usepackage[colorlinks=true, citecolor=red, linkcolor=blue, urlcolor=blue]{hyperref}
\begin{document}
\title{Cu-Doped KCl folded and unfolded band structure and optical properties studied by DFT calculations}
\author{Jose Luis Cabellos$^1$}\email[email:]{sollebac@gmail.com}
\author{C\'esar Castillo-Quevedo$^2$}
\author{Raul Aceves$^1$}
\author{Roberto Nunez-Gonzalez$^3$}
\author{Alvaro Posada-Amarillas$^1$}
\affiliation{$^1$Departamento de Investigaci\'on en F\'isica, Universidad de Sonora, Blvd. Luis Encinas y Rosales S/N, 83000 Hermosillo, Sonora, M\'exico}
\affiliation{$^2$Departamento de Fundamentos del Conocimiento,
  Centro Universitario del Norte, Universidad de Guadalajara,
Carretera Federal No. 23, Km. 191, C.P. 46200, Colotl\'an, Jalisco, M\'exico}
\affiliation{$^3$Departamento de Matem\'aticas, Universidad de Sonora, Blvd. Luis Encinas y Rosales S/N, 83000 Hermosillo, Sonora, M\'exico.}
\date{\today}
\affiliation{\textcolor{black}{$^*$corresponding author: sollebac@gmail.com}}
\begin{abstract}
  We computed the optical properties and the folded and unfolded band structure of Cu-doped KCl crystals. The calculations use the plane-wave pseudo-potential approach implemented in the ABINIT~\cite{abinit,abinit1} electronic structure package within the first-principles density-functional theory framework.  Cu substitution into pristine KCl crystals requires calculation by the supercell (SC) method from a theoretical perspective. This procedure shrinks the Brillouin zone, resulting in a folded band structure that is difficult to interpret. To solve this problem and gain insight into the effect of cuprous ion (Cu$^+$) on electronic properties; We unfolded the band structure of SC KCl:Cu to directly compare with the band structure of the primitive cell (PC) of pristine KCl. To understand the effect of Cu substitution on optical absorption, we calculated the imaginary part of the dielectric function of KCl:Cu through a sum-over-states formalism and broke it down into different band contributions by partially making an iterated cumulative sum (ICS) of selected valence and conduction bands. As a result, we identified those interband transitions that give rise to the absorption peaks due to the Cu$^+$ ion. These transitions include valence and conduction bands formed by the Cu-3d and Cu-4s electronic states.  To investigate the effects of doping position, we consider different doping positions, where the Cu dopant occupies all the substitutional sites replacing host K cations. Our results indicate that the doping position's effects give rise to two octahedral shapes in the geometric structure.  The distorted-twisted octahedral square bipyramidal geometric-shape induces a difference in the \emph{crystal field splitting energy} compared to that of the perfect octahedral square bipyramidal geometric-shape. Therefore, the optical anisotropy's origin and the change in bandgap come from the difference in the \emph{crystal field splitting energy} between the distorted-twisted and the perfect octahedral square bipyramidal geometric-shapes.\\
\end{abstract}
\maketitle
\section{Introduction}
Alkali halide (AH) crystals are solids of great importance from theoretical and experimental points of view. They are of great research interest in solid-state physics,\cite{bhandari} mainly due to their high stability.~\cite{baldochi} Pure AH crystals are relatively easy to produce in large quantities. They possess high melting points, varying from 600 to 1000 °C,~\cite{sirdeshmukh} are poor conductors of heat,~\cite{baldochi} and have strong miscibility in polar media.~\cite{gopikrishnan} They are also the most ionic of all crystal compounds~\cite{baldochi} that consist of ions bound together by electrostatic attraction, making them good candidates for studying other systems.~\cite{chaofan} The AH crystals have a large energy gap in the order of 8–10 eV, making them useful for the development of laser optical components as optical transmission windows in the ultraviolet (UV) to infrared (IR) ranges of the electromagnetic spectrum,~\cite{sirdeshmukh,zjchen} among other optical applications. AH crystals, either pure or doped, are also employed as neutron monochromators.~\cite{sirdeshmukh}
Potassium chloride (KCl) is an inorganic salt with properties similar to those of common salt. (NaCl)~\cite{katica} KCl is employed in optical windows for laser applications~\cite{palik} and as a scintillator or in dosimetry by adding impurities.~\cite{feridoun,silviu,shiehpour} At the end of the 1960s, several experimental and theoretical studies on the optical properties of KCl and other AH crystals were presented.~\cite{tomiki,baldini,baldini2,said,philipp,blechschmidt,kondo} Recently, theoretical predictions on novel KCl phases under pressure have been proposed.~\cite{bahattin,zjchen} Several studies have addressed doping of KCl with monovalent cations, e.g., Ga$^+$, In$^+$, and Tl$^+$, to investigate distortion trends as a function of impurity.~\cite{aguado} Furthermore, recently published works show an enhancement in the optical properties of KCl crystals when doped with ZnO.~\cite{shiehpour} In other previous works, the optical properties of KCl crystals doped with Sb$_2$O$_3$ nanocrystals were investigated, analyzing the doping effects of the Sb$_2$O$_3$ nano-semiconductor on the optical properties of KCl crystals.~\cite{bouhdjer} The optical properties of copper ion (Cu$^+$)-doped KCl and other AH crystals have also been studied due to their outstanding UV-light absorption properties,~\cite{myasnikova2,joseph,winter1987} a characteristic that makes them excellent phosphor materials for many technological applications.~\cite{preto2017,preto2016,nunez}  In the Cu$^+$-doped KCl, several absorption bands in the UV region have been observed, with a strong dependence on temperature.~\cite{goldberg1987,myasnikova2} These bands have been attributed to 3d$^{10}\rightarrow$3d$^9$4s$^1$ transition states of Cu$^+$ ion. Theoretical calculations of the oscillator strength of Cu$^+$ 3d$^{10}\rightarrow$3d$^9$4s$^1$ transitions in NaF hosts were performed to verify this conclusion.~\cite{elmar2012}  Recently, density functional theory (DFT) calculation of the electronic and optical properties of LiF:Cu$^+$ was carried out, employing the materials modelling code based on a quantum mechanical description of electrons (CASTEP code) within the pseudopotential approximation and reporting the formation of a band in the middle of the electronic bandgap due to copper (Cu).~\cite{sun2012}   Despite the remarkable efforts that have been made to elucidate the electronic structure and origin of the absorption and emission spectra of the Cu$^+$ ion embedded in KCl and other AH crystals, this has still not been achieved, and further theoretical studies with different models and methodologies are needed.
In this paper, we report the results of DFT calculations of the band structure, density of states, and optical properties of the Cu$^+$ ion embedded in KCl. To study and identify the effects that impurities have on its electronic structure, we unfolded the band structure of KCl:Cu and compared it with that of pristine KCl. The band-to-band contribution to the optical spectrum is examined by partially summing selected bands using \emph{the sum-over-states formalism},\cite{cabellos}  which is employed in the evaluation of the imaginary part of the linear electric susceptibility tensor. The transitions between valence and conduction bands that are responsible for the effects of Cu substitution on the absorption of KCl are thus identified. To investigate the effects of doping position, we consider different doping positions, where the Cu dopant occupies all the substitutional sites replacing host K cations. Our results indicate that the doping position's effects give rise to two octahedral shapes in the geometric structure. The remainder of this paper is organized as follows: Section 2 briefly outlines the theory and provides the computational details. Section 3 presents a detailed discussion of the results of the unfolding and folding band structure, density of states, and band-to-band contribution to the optical response of pristine KCl and doped KCl:Cu system. Conclusions are given in Section 4.

\section{Theoretical methods and computational details}
\begin{figure}[ht!]
  \begin{center}
  \setcounter{subfigure}{0} 
  \subfloat[][]{\includegraphics[scale=0.3]{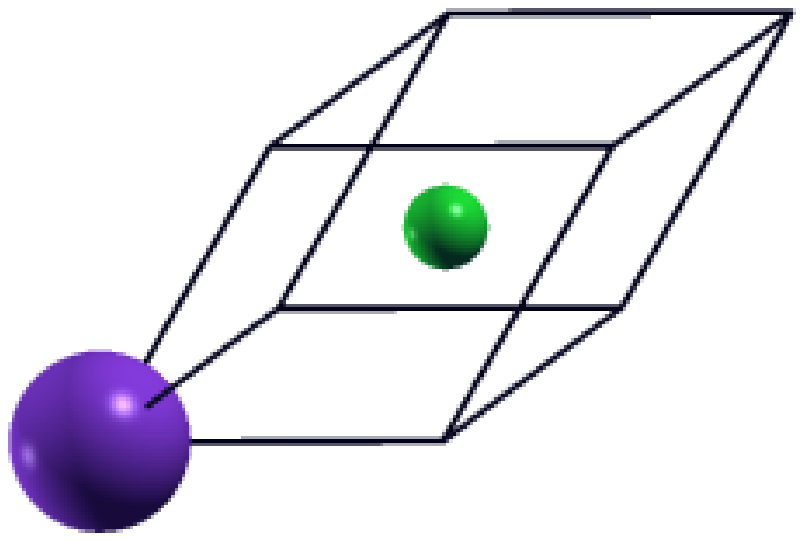}}\hspace*{\fill} 
  \subfloat[][]{\includegraphics[scale=0.3]{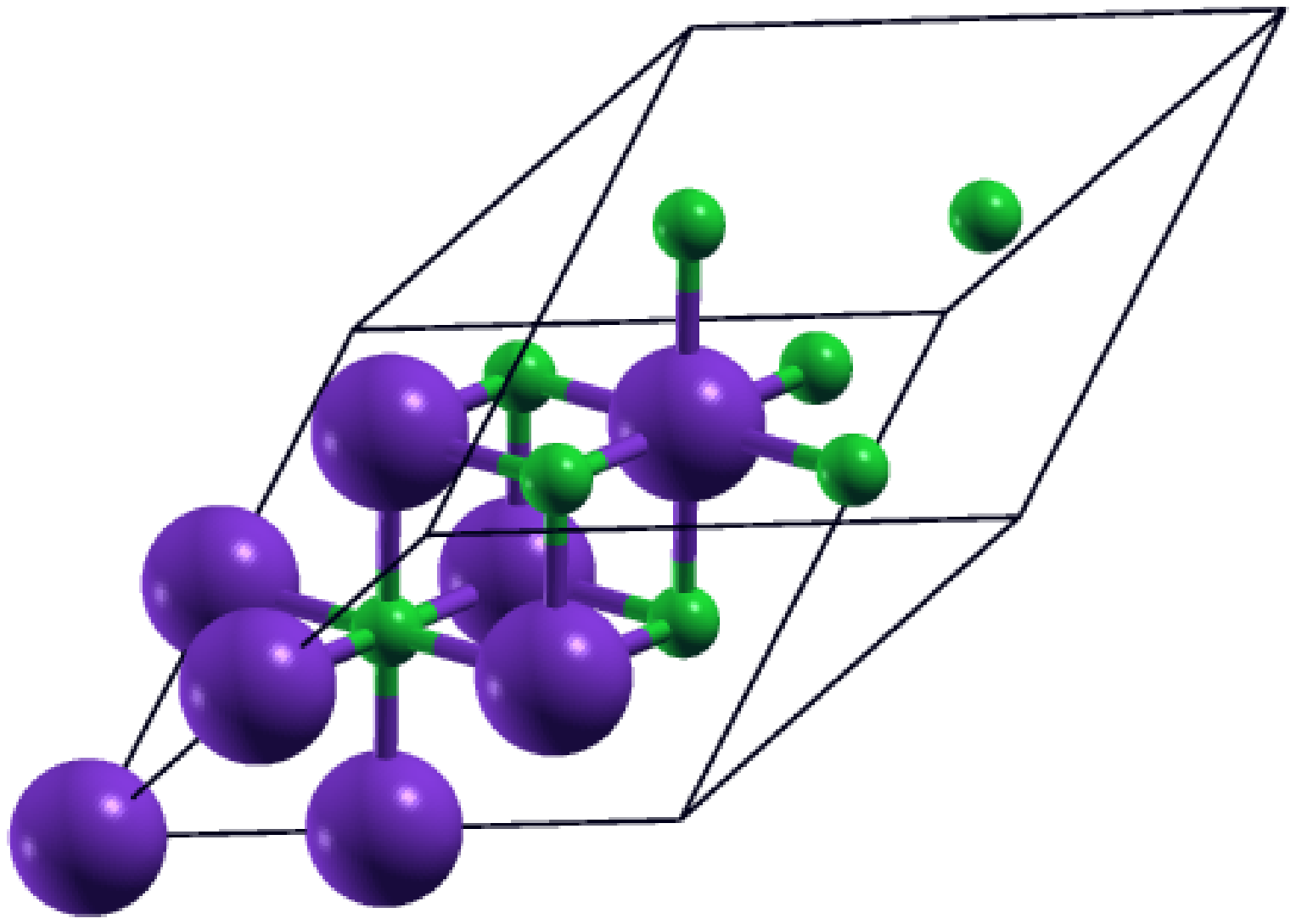}}\hspace*{\fill}
  \caption{(Color online.) (a) The pristine FCC PC of the KCl with two atoms at the basis.
    (b) The pristine FCC SC 2x2x2 of the KCl with 16 atoms at the basis
    The violet, green, and copper colored spheres
    represent the Potassium, Chlorine and Copper atoms respectively. The unit
    cell size as well as the atoms size are depicted proportional.}
  \label{cells}
  \end{center}
\end{figure}
\begin{figure}[ht!]
\centering
\includegraphics[scale=0.65]{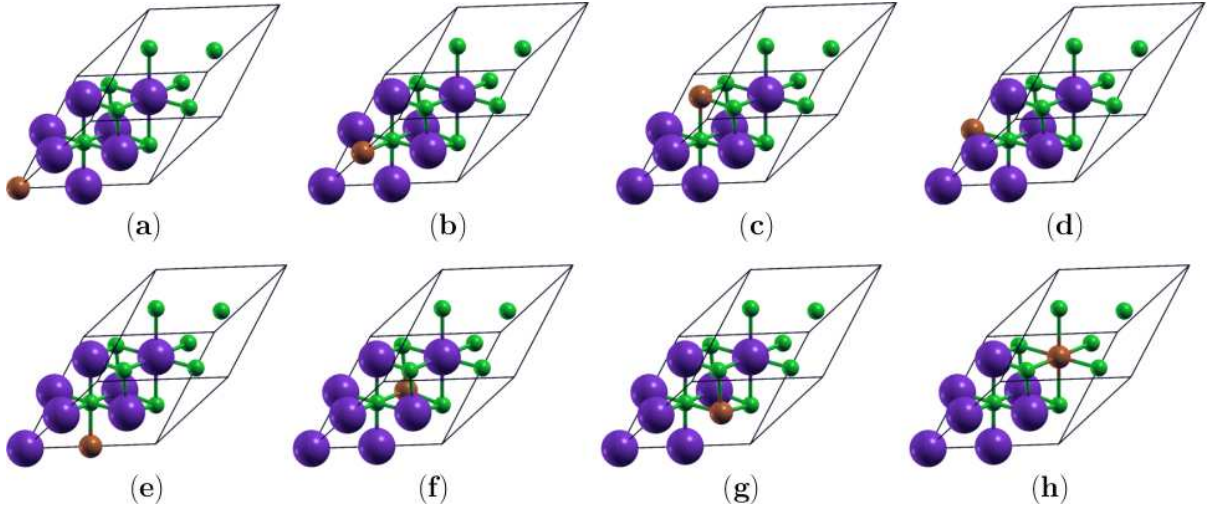}
\caption{(Color online.) From (a) to (h) supercell of KCl doped with Cu atom with 16 atoms at the base. We considerer the dopants occupy only the substitutional sites replacing all host K cations. The violet-, green-, and copper-colored spheres represent the potassium, chlorine, and copper atoms.}
\label{doping}
\end{figure}

KCl is an ionic crystal with a rock salt structure (NaCl). The face-centered cubic (FCC) crystalline structure phase of KCl belongs to the space group (Hermann–Mauguin) Fm3m (number 225) and point group (Hermann–Mauguin) m3m. In this structure, an octahedron of six counterions surrounds each ion; hence, the coordination number is six for both the anion, Cl−, and cation, K$^+$. The experimental lattice constant is 6.29 \AA.~\cite{walker,schwabegger,chun-gang,sirdeshmukh} Figure~\ref{cells}a shows the FCC primitive cell (PC) of pristine KCl, with two atoms at the base; the cation K$^+$ is located at the [0,0,0]
position, and the anion Cl$^−$ is located at the [1/2,1/2,1/2] position, for which the Brillouin zone (BZ) is a truncated octahedron [36,37],~\cite{aroyo,setyawan2010} and its primitive vectors are [0, 1/2,1/2], [1/2,0,1/2], and [1/2,1/2,0]. Figure~\ref{cells}b shows a larger FCC pristine system that contains eight times as many atoms as in the PC, so the number of atoms is 16 (8 Cl$^−$ and 8 K$^+$). This cubic phase was obtained by accurately multiplying the PC lattice constants of KCl by $2\times 2\times 2$. As a result, the lattice vector length became twice that of the original unit cell of the PC and has the same primitive vectors. This system allows us to make Cu substitutions, doping the KCl structure. Moreover, to gain insight into the effect of Cu dopant in electronic and optical properties of pure KCl, we consider the exploration of the other eight possible sites that the Cu atom replaces the host cation K shown in Figure.~\ref{doping}. We computed the geometric, electronic, and optical properties for those eight systems. The doping position affects the bandgap, and optical properties of the KCl:Cu system. Interesting, it induces an anisotropic nanostructure, as a consequence, induces optical anisotropy. 
\begin{table}[ht!]
  \caption{Parameters employed in the present study’s calculation.}
  \label{geom}
\centering
\begin{tabular}{lccc}
  \toprule
  \hline
  \hline
\textbf{Unit cell}	& \textbf{Number of k-points}	& \textbf{Size grid} & \textbf{ecut-off (Ha)}  \\
\hline
Fig.\ref{cells}a~KCl		     &  3864			& 55$\times$55$\times$55 & 25\\
Fig.\ref{cells}b~K$_8$Cl$_8$      &  2920			& 19$\times$19$\times$19 & 25\\
Fig.\ref{doping}a~K$_7$Cl$_8$:Cu   &  2920			& 19$\times$19$\times$19 & 25\\
Fig.\ref{doping}b~K$_7$Cl$_8$:Cu   &  2920			& 19$\times$19$\times$19 & 25\\
Fig.\ref{doping}c~K$_7$Cl$_8$:Cu   &  2920			& 19$\times$19$\times$19 & 25\\
Fig.\ref{doping}d~K$_7$Cl$_8$:Cu   &  2920			& 19$\times$19$\times$19 & 25\\
Fig.\ref{doping}e~K$_7$Cl$_8$:Cu   &  2920			& 19$\times$19$\times$19 & 25\\
Fig.\ref{doping}f~K$_7$Cl$_8$:Cu   &  2920			& 19$\times$19$\times$19 & 25\\
Fig.\ref{doping}g~K$_7$Cl$_8$:Cu   &  2920			& 19$\times$19$\times$19 & 25\\
Fig.\ref{doping}h~K$_7$Cl$_8$:Cu   &  2920			& 19$\times$19$\times$19 & 25\\
\hline
\end{tabular}
\end{table}
In the following, we studied the eight systems focusing on the doped isotropic system of Figure.~\ref{doping}a, and the doped anisotropic system shows in Figure.~\ref{doping}b.  With these doped systems, we calculate the optical and electronic properties, and the electronic and optical properties of the pristine supercell SC/PC work as a reference system to observe the effect of the Cu$^+$ ion. Particular attention should be paid to the SC method of calculation. This method conveniently allows the folding of the bands into the smaller SC BZ that gives rise to a complicated band structure for which it is difficult to determine whether the bandgap is direct or indirect.~\cite{cartoixa,popescu} Moreover, other physical quantities, such as carrier mobility,~\cite{maspero} require knowledge of the PC band structure. Several authors have already recently developed theories and methods with diverse approximations to unfold the SC band structure into a PC band structure, with successful results.~\cite{dargam,boykin,popescu,popescu2,chen,lee,ku,allen,rubel,medeiros,medeiros2,huang,kosugi,sara} In this study, the unfolding of the SC BZ into the PC BZ was performed using the BandUP code,~\cite{medeiros,medeiros2} which is based on the evaluation of the spectral weight given by Eq.~\ref{spectral} 
\begin{equation}
 \displaystyle 
 P_{mK}(\vec{k})=\sum_n \vert \langle\psi_{nk}^{PC} \vert \psi_{mK}^{SC}\rangle \vert^2
 \label{spectral}
\end{equation}
where $P_{mK}(\vec{k})$ is the spectral weight, and is a reciprocal ${\bf{k}}$-point, the $\vert\psi_{nk}^{PC}\rangle $ and $\vert\psi_{mK}^{SC}\rangle$ are the eigenstates of the PC and SC respectively, and it gives the probability of an eigenstate of Hamiltonian, in the SC representation, to have the same character as a PC state. This unfolding procedure of the SC BZ has been successfully employed in several systems, such as SiGe nanowires,~\cite{cartoixa}  graphene,~\cite{dombrowski,warmuth} and ternary alloys.~\cite{arash}
The optical absorption spectrum is determined by the dielectric function,~\cite{molina2013}, which is related to the electric susceptibility, ($\mathfrak{Im}[\chi_1^{ab}(-\omega,\omega)]$)~\cite{sipe}
by Eq.~\ref{e1}
\begin{equation}
\displaystyle
\epsilon^{ab}(\omega)=1+4\pi\chi_1^{ab}(-\omega;\omega)
\label{e1}
\end{equation}
where $ab$ letters are the axes $x$, $y$ and $z$ of an orthogonal coordinate system and $\omega$ is the frequency of the light. 
The imaginary part of electric susceptibility it is given in Eq.~\ref{chi1}.
\begin{equation}
\displaystyle
\mathfrak{Im}[\chi_1^{ab}(-\omega,\omega)]=\frac{e^2}{\hbar}\int\frac{d^{3}k}{8\pi^3}\sum_{n\neq m} 
{f_{nm} r_{nm}^{a}({\bf{k}})r_{mn}^{b}({\bf{k}})}
\delta(\omega_{mn}-\omega) 
\label{chi1}
\end{equation}
In Equation~\ref{chi1}, $e$ and $\hbar$ are the elementary charge and Planck’s constant, respectively;  and $n$, and $m$ are the occupied initial and unoccupied final states; $f_{nm}$ is the Fermi occupation factor, which is zero or unity for cold semiconductors; 
$\omega_{mn}({\bf{k}})$ are the frequency differences, where $\omega_{m}({\bf{k}})$, and $\omega_{n}({\bf{k}})$  are the DFT energies of bands $n$, and $m$ at wave vector ${\bf{k}}$; and $r_{nm}^{a}({\bf{k}})$ are the matrix position elements of bands $m$ and $n$ at vector ${\bf{k}}$.
Following the procedure of a previous work,~\cite{cabellos}
\begin{table}[ht!]
  \caption{Lattice constant and bandgap obtained in the DFT calculation for all systems where Cu dopant occupies substitutions sites replacing host K cations. }
  \label{geom2}
\centering
\begin{tabular}{lccc}
  \toprule
  \hline
  \hline
\textbf{Unit cell}	& \textbf{Type of cell} & \textbf{Lattice-constant~(\AA) } &\textbf{DFT-bandgap~(eV)}  \\
\hline
Fig.\ref{cells}a~KCl		   & Face-Centered Cubic primitive  & 4.507  & 5.07\\
Fig.\ref{cells}b~K$_8$Cl$_8$       &  Face-Centered Cubic supercell & 9.014   & 5.07\\
Fig.\ref{doping}a~K$_7$Cl$_8$:Cu   &  Face-Centered Cubic supercell & 8.390  & 2.20\\
Fig.\ref{doping}b~K$_7$Cl$_8$:Cu   &  Face-Centered Cubic supercell & 8.480  & 3.20\\
Fig.\ref{doping}c~K$_7$Cl$_8$:Cu   &  Face-Centered Cubic supercell & 8.480  & 3.20\\
Fig.\ref{doping}d~K$_7$Cl$_8$:Cu   &  Face-Centered Cubic supercell & 8.580  & 3.20\\
Fig.\ref{doping}e~K$_7$Cl$_8$:Cu   &  Face-Centered Cubic supercell & 8.480  & 3.20\\
Fig.\ref{doping}f~K$_7$Cl$_8$:Cu   &  Face-Centered Cubic supercell & 8.480  & 3.20\\
Fig.\ref{doping}g~K$_7$Cl$_8$:Cu   &  Face-Centered Cubic supercell & 8.480  & 3.20\\
Fig.\ref{doping}h~K$_7$Cl$_8$:Cu   &  Face-Centered Cubic supercell & 8.394  & 3.20\\
Fig.\ref{cubic}a~K$_4$Cl$_4$       &  Cubic supercell               & 6.378  & 5.07\\
Fig.\ref{cubic}a~K$_4$Cl$_4$:Cu    &  Cubic supercell               & 6.180   & 0.81\\
\hline
\end{tabular}
\end{table}
in this study, we computed the expression given in Equation~\ref{chi1}. The peaks, presented in the $\mathfrak{Im}[\chi_1^{ab}(-\omega,\omega)]$ part, are due to the direct interband optical transitions between the valence and conduction bands, which can be identified from the critical points of the band structure. Moreover, in this study, the bands that contribute significantly to the $\mathfrak{Im}[\chi_1^{ab}(-\omega,\omega)]$  part are identified by breaking down the $\mathfrak{Im}[\chi_1^{ab}(-\omega,\omega)]$ part into different band contributions.~\cite{salazar,lee2004} In addition, we considered the scissor correction,~\cite{nastos,cabellos}  the approximation of which is obtained by a rigid shift of the DFT energies to an experimental value.\cite{stahrenberg} Therefore, the spectrum of $\mathfrak{Im}[\chi_1^{ab}(-\omega,\omega)]$   is rigidly shifted along the energy axis without changing the spectrum shape.~\cite{cabellos,nastos} The real $\mathfrak{Re}[\chi_1^{ab}(-\omega,\omega)]$ part  was computed using Kramers-Kronig relations. Here, the indirect transitions are neglected because they have little contribution~\cite{okoye} to the $\mathfrak{Im}[\chi_1^{ab}(-\omega,\omega)]$ part and the spin–orbit, local field, and electron-hole effects are also neglected.~\cite{leitsmann,cabellos} The inclusion of these effects is beyond the scope of this study.
The electronic and optical properties are calculated using the SC approach in the DFT framework, as is implemented in the freely available ABINIT software code.~\cite{abinit,abinit1}. We chose the optimized norm-conserving Vanderbilt (ONCV) pseudopotentials~\cite{hamann2013} taken from the ONCVPSP-PBE-PD v0.4 library, following reference.~\cite{abinit}; These were validated against all-electron calculations and found to perform well. Among these
 pseudopotentials the 3s$^2$3p$^6$3d$^{10}$4s$^1$, 3s$^2$3p$^6$4s$^1$, and 3s$^2$ 3p$^5$ electrons of the Cu, K, and Cl atoms,
respectively, are treated as valence states. We employed the Perdew-Burke-Ernzerhof (PBE) general gradient approximation functional to calculate the exchange-correlation energy.~\cite{perdew1996,perdew1992} The wavefunctions were expanded in a plane-wave basis set and checked for convergence by applying a kinetic cutoff energy of 25 Ha. The Monkhorst-Pack scheme~\cite{monk} was used to sample the irreducible Brillouin zone (IBZ). The IBZ integration was performed by employing the tetrahedron method.~\cite{cabellos,PhysRevB.84.195326,PhysRevB.80.245204,PhysRevB.85.165324,girlanda} Table~\ref{geom} shows the number of the chosen k-points, cutoff energy, and k-point mesh corresponding to all systems, to assure convergence of the total energy and forces as well as the optical properties. The total energy of the self-consistent field (SCF) procedure was set to a value of 5$\times$10$^7$ eV/atom. All the atomic positions of the studied systems were relaxed until the Hellmann-Feynman forces on each atom were lower than 20 meV/\AA. The resulting atomic structures were used for all the calculations presented in this study.

\section{Results and Discussion}
\subsection{Optimized Lattice Constant}
\begin{figure}[ht!]
\centering
\includegraphics[scale=0.65]{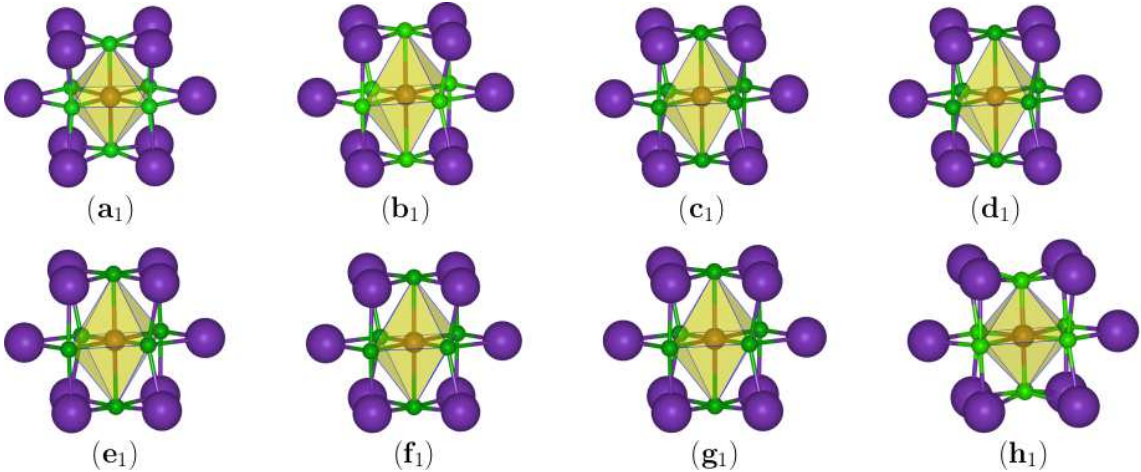}
\caption{(Color online.) The panel (a$_1$) and panel (h$_1$)  show the octahedral square bipyramidal shape.  In this geometry,  the six Cl atoms are symmetrically arranged around a central Cu atom.  The optimized bond length between the Cu atom and the six surrounding Cl atoms is 2.651~\AA.~The panel (b$_1$-g$_1$)  shows slightly twisted octahedra square bipyramidal shape. In this octahedrons, there are four optimized bond lengths of Cu-Cl of 3.146~\AA~ and two of 2.151~\AA.~   The elongation of four bounds, in the slightly twisted octahedron square bipyramidal shape, could be related to a \emph{Jahn-Teller effect}, and  the distorted system could be more stable than the undistorted one. The violet-, green-, and copper-colored spheres represent the potassium, chlorine, and copper atoms, respectively, and the octahedra are a yellow color}
\label{octa}
\end{figure}
We computed the PC KCl lattice constant through minimization of the total energy by employing the parameters shown in Table~\ref{geom}. The obtained result was 4.507~\AA~for the FCC PC system, which is in agreement with the experimental lattice constant of 6.29~\AA~\cite{walker,schwabegger,chun-gang,sirdeshmukh} —larger by 1.22{\%} and is slightly shorter by 0.3{\%}—compared with other theoretical calculations that obtained 4.522~\AA.~\cite{persson} It is more similar to other calculations that used an all-electron scheme~\cite{nunez} (see Table~\ref{geom2}). The relation between cells FCC-PC and cubic SC with eight atoms basis (Figure~\ref{cubic}) is a factor of $\sqrt{2}$. ($4.507\times \sqrt{2}  = 6.37$). In the Appendix~\ref{appendix:a} Figure~\ref{cubic} shows the pure KCl SC on the basis of eight atoms. 
\begin{figure}[ht!]
\centering
\includegraphics[scale=0.6]{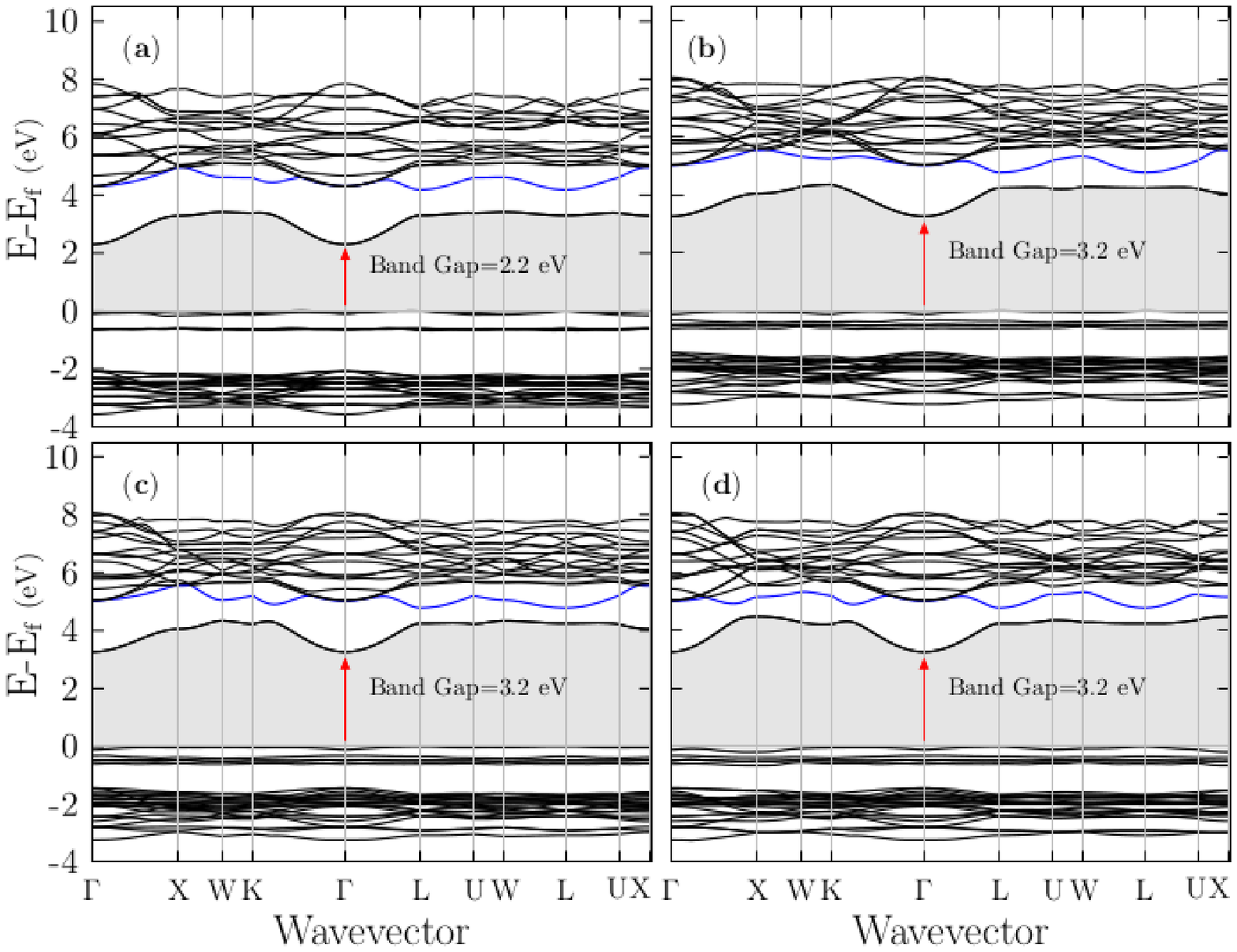}
\includegraphics[scale=0.6]{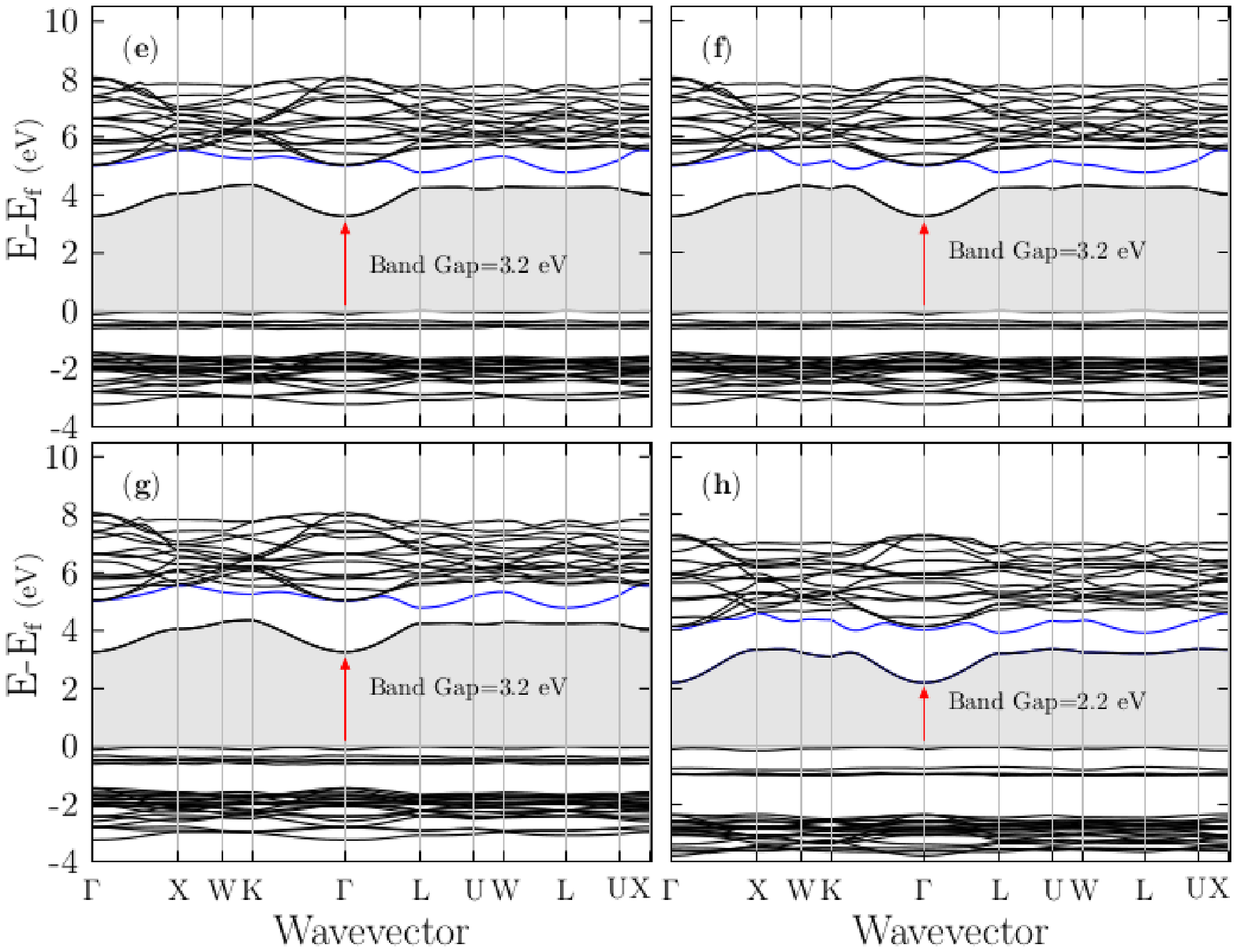}
\caption{(Color online.) The  panels (a,b,c,d,e,f,g,h) shows the calculated Kohn-Sham folded band structure of SC K$_7$Cl$_8$:Cu for the cases of Figure~\ref{doping}(a,b,c,d,e,f,g,h)  respectively and along with the same high-symmetry directions of the BZ used in the PC. The lowest direct Kohn-Sham bandgap, indicated by the red arrow, is  2.2 eV at the point for the case in Figure~\ref{doping}a and 3.2 eV. for the other three cases and indicated by the red arrow.  The solid blue line depicts the second conduction band; notice there is a little change in its shape.  The first conduction band shape almost keeps unchanged. }
\label{bands1}
\end{figure}
The FCC supercell of the pristine system K$_8$Cl$_8$ is eight times the volume of the primitive cell of KCl. In Table~\ref{geom2}, the lattice constant of the SC two times greater than the lattice constant of the PC. Hereafter, we focus on the symmetric octahedron system shown in the Figure~\ref{doping}a.  After the optimization of the K$_7$Cl$_8$:Cu, the resulting unit cell is FCC with its three optimized lattice parameters measuring 8.390~\AA,~and its three lattice angles being 60$^{\circ}$.  The optimized volume of the pristine K$_8$Cl$_8$ is 472.0~\AA$^3$,~ and the volume of K$_7$Cl$_8$:Cu is 417.7.~\AA$^3$~ The SC volume of the doped system is 11.5{\%} smaller, which is evidently due to copper doping (in Figure~\ref{cells}, this can be seen in the proportions of the unit cells). For this system, the calculated density of Cu based on its atomic weight and the volume of the unit cell with an 8.390~\AA~ edge length is 0.25 g/cm$^3$. The unit cell has five non-equivalent atoms, and the distances between the Cu atom and the ions surrounding it are shorter by approximately 0.53~\AA.~ The optimized bond length of K-Cu is 2.651 ~\AA~ compared to the bond length of K-Cl atoms (3.187 ~\AA~) in the pristine system. In Appendix~\ref{appendix:b}, Table~\ref{list} is given the list of non-equivalent atoms in the unit cell; Appendix~\ref{appendix:b}, Table~\ref{list2} is given the list of atoms in the unit cell, and  Appendix~\ref{appendix:b}, Table~\ref{list3} is given the three lattice vectors, of the KCl:Cu system of Figure~\ref{doping}a. The exploration of the other possible sites after optimization indicate that there are two optimized lattice parameters shown in Table~\ref{geom2} and that there is a relative displacement of doped and host atoms as a function of dopant occupying the site. The displacement at each lattice site depends on the Cu atom occupying that site and the which kind of atoms are closer to the Cu atom. To better appreciate, Figure~\ref{octa} shows the coordination polyhedron for each system. It is clear that, if we take the polyhedron of the pure KCl as a reference, the eight polyhedra of the Cu doped systems are distorted. On a closer inspection of Figure\ref{octa}, we can distinguish that there are two shapes of coordinate polyhedra. Figure~\ref{octa}a shows regular spherical polyhedron, where the optimized bond length between the Cu atom and the six surrounding Cl atoms is 2.651~\AA.~Figure~\ref{octa}b shows a prolate polyhedron, where there are two types of optimized bond length among the Cu, and six surrounding Cl atoms, four optimized bound lengths of Cu-Cl are 3.146~\AA;~ the other two are 2.151~\AA.~The rest of the polyhedra are of one type or another. The Cu dopant occupying different sites of the KCl gives rise to two different geometrical systems that cause significant changes in their electronic and optical properties of the doped KCl.

\section{Electronic Band Structure and Density of States}
\begin{figure}[ht!]
\centering
\includegraphics[scale=0.65]{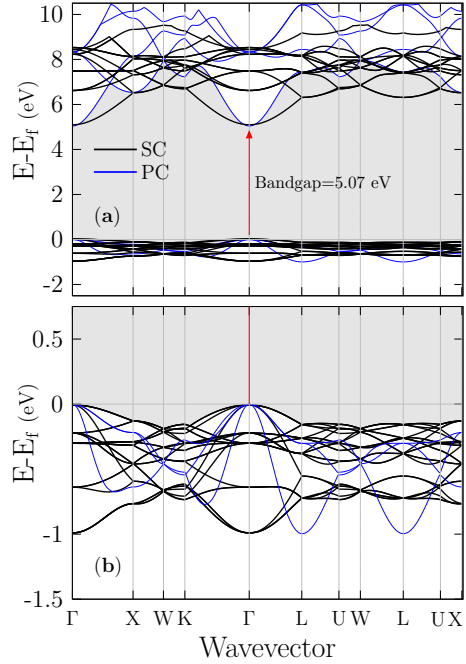}
\caption{(Color online.) The upper panel (a) shows the band structure of the primitive cell (PC) for KCl. (blue lines) and the folded band structure of the supercell (SC) of pristine K$_8$Cl$_8$ (black lines) along with the same high-symmetry directions of the Brillouin zone (BZ) of the PC. The lower panel (b) shows a zoomed view in the region -1.5 to 0.75 eV. The shortest direct Kohn-Sham bandgap indicated by the red arrow is ~5.07~eV at the $\Gamma$ point.  The blue lines correspond to the PC band structure, and the black lines correspond to the SC band structure. Notice, the change in band dispersion in spite,  they are computed in the same materiaL. Figure~\ref{scissors} in the Appendix~\ref{appendix:b} shown the PC band structure with applied scissors correction.}
\label{bands2}
\end{figure}
The band structure of the pristine KCl has been reported previously.~\cite{chun-gang,nunez,persson} For comparison, in later sections with the eight doped system (Figure~\ref{doping}),  
we show in Figure~\ref{bands1} the band structure  of all eight cases of the Cu-doped  KCl computed along high-symmetry directions in the PC BZ, from the BZ center  with the coordinates (0,0,0) to the X point (0,0.5,0.5), W point (0.25,0.5,0.75), K point (0.375,0.375,0.75),  point (0.0,0.0,0.0), L point (0.0,0.0,0.5), U point (0.0,0.375,0.625), W point (0.25,0.5,0.75), L point (0.0, 0.0,0.5), U point (0.0,0.375,0.625), and X point (0.0 0.5 0.5) in units of , where a, b, and c are the lattice constants shown in Table~\ref{geom2}. Figure~\ref{bands2}a shows the unfolded band structure of the pristine SC K$_8$Cl$_8$, Fig.\ref{cells}b, and black lines, superimposed to the band structure of pristine PC, Fig.\ref{cells}a), blue lines.  Figure~\ref{bands2}b shows a zoomed view in the region -1.5 to 0.75 eV. The lowest direct Kohn-Sham band gap of the KCl obtained in both calculations is ~5.07 eV, at the $\Gamma$ point, as indicated by the red arrow in Figure~\ref{bands2}ab (displayed in Table~\ref{geom2}) and in agreement with previous calculations.~\cite{nunez,persson}
\begin{figure}[ht!]
\centering
\includegraphics[scale=0.65]{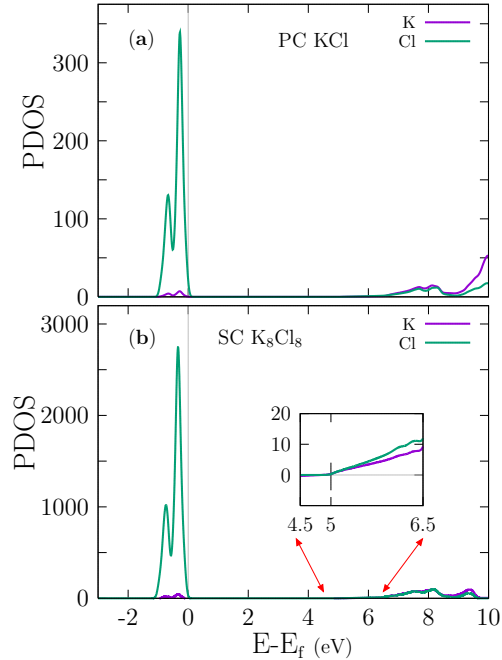}
\caption{(Color online.) The upper panel (a) shows the Kohn-Sham projected density of states (PDOS) for the  pristine PC KCl. The lower panel (b) shows the PDOS for the pristine SC of K$_8$Cl$_8$. The inset in the lower panel (b) of  Figure 3 shows a zoomed view in the range 4.5 to 6.5 eV energy axes, clearly shows that density of states start to raise at energy range of 5 eV, as we expected. Notice the main difference: the density is higher in the SC than in the PC system. The color code is violet for K and green for Cl atoms. In both plots, the Cl-3p states dominate at the Fermi level.} 
\label{dos1}
\end{figure}
In these calculations, the bandgap is underestimated by ~3.53 eV compared with the experimental value of 8.6 eV of the pure KCl, and other recently reported all-electron DFT calculations using the modified Becke-Johnson approximation.~\cite{nunez} To correct the bandgap that was underestimated by DFT,~\cite{jones} we applied a scissor correction of 3.53 eV to the conduction bands, which shifted them to the experimental value. Following and comparing the results presented in Figure~\ref{bands2}ab, it can be seen that the principal difference is the band dispersion in the K-$\Gamma$-L direction, although we consider the same crystal. This difference is due to SC shrink of the BZ; thus, one should not wrongly conclude something that depends on the dispersion of the band structure (e.g., the effective mass) using an SC band structure.
\begin{figure}[ht!]
\centering
\includegraphics[scale=0.65]{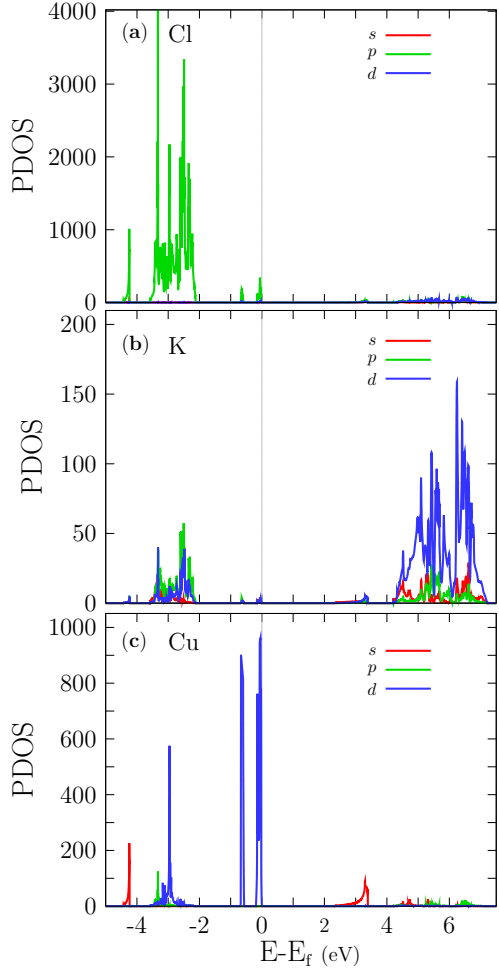}
\caption{(Color online.) PDOS for system shown in Figure~\ref{doping}a. The upper panel (a): Kohn-Sham PDOS for the Cl atom. Middle panel (b): Kohn-Sham PDOS for the K atom. The lower panel (c): PDOS for the Cu atom. The Cu-3d states dominate at the Fermi level with a sharply peaked density, and at 2.2 eV, the Cu-4s states dominate. For all atoms, s-, p-, and d-states are identified by red, green and blue lines, respectively.}
\label{dos2}
\end{figure}
To elucidate the nature of the electronic band structure, we calculated the PDOS as a function of energy for the pristine PC KCl (Figure~\ref{dos1}a), SC K$_8$Cl$_8$ (Figure ~\ref{dos1}b), and doped SC K$_7$Cl$_8$:Cu (Figure 4a–c) structures. In those plots, the Fermi energy is shifted to zero. For the pristine PC and SC, our results are in good agreement with those of previous studies; notice that the density of PDOS for the SC is higher than that for the PC. The examination of the PDOS shows that the width of valence bands is approximately 1 eV, which is in agreement with the folded band structure (black lines) shown in Figure~\ref{bands1}a . The highest valence bands are mainly formed by the Cl-3p states, which are in line with previous studies [35]. Just slightly below the Fermi level, there are tiny contributions of K-3p states, as can be seen from the PDOS plot in Figure.~\ref{dos1}  Thus, there is a slight admixture (not an overlap) between the Cl-3p states and a few K-3p states, as expected for an ionic compound. Meanwhile, at approximately 5 eV, the bottom conduction bands have contributions from both the K-4s and Cl-4s states.
\section{Folded band structure of the doped K$_7$Cl$_8$:Cu}
\begin{figure}[ht!]
\centering
\includegraphics[scale=0.65]{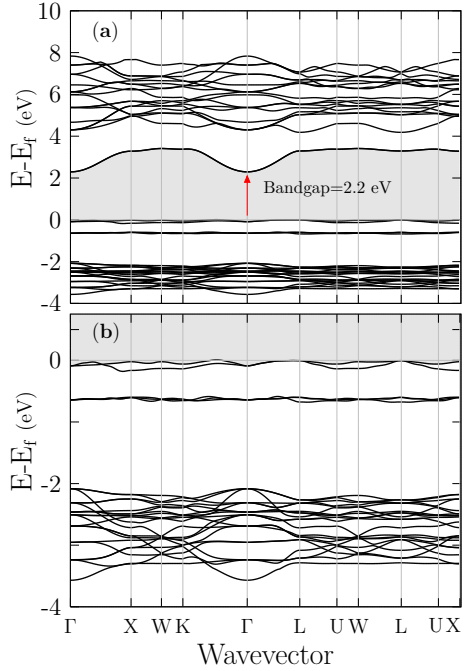}
\caption{(Color online.) The upper panel (a) shows the calculated Kohn-Sham folded band structure of SC K$_7$Cl$_8$:Cu for the symmetric case of Figure ~\ref{doping}a, and along the same high-symmetry directions of the BZ used in the PC. The lower panel (b) shows a zoomed region in the energy range of 4.0 to 0.5 eV. The lowest indirect Kohn-Sham bandgap, indicated by the red arrow, is 2.2 eV at $\Gamma$ the point.}
\label{bands3}
\end{figure}

\begin{figure}[ht!]
\centering
\includegraphics[scale=0.65]{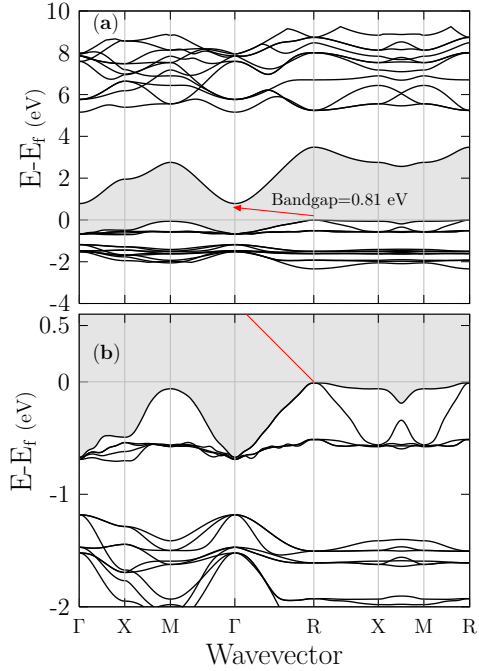}
\caption{(Color online.) The upper panel (a) shows the folded bandstructure of the Cu doped supercell  for KCl (CuK$_3$Cl$_4$)  along the high-symmetry directions of the Brillouin zone (BZ) for a cubic cell. The lower panel (b) shows a zoomed view in the region  -2 to 0.75 eV. The shortest indirect Kohn-Sham bandgap indicated by the red arrow is 0.80 eV between and R points. Notice that the bandgap character changes from direct to indirect as the Cu concentration increases.}
\label{bands4}
\end{figure}

\begin{figure}[ht!]
\centering
\includegraphics[scale=0.6]{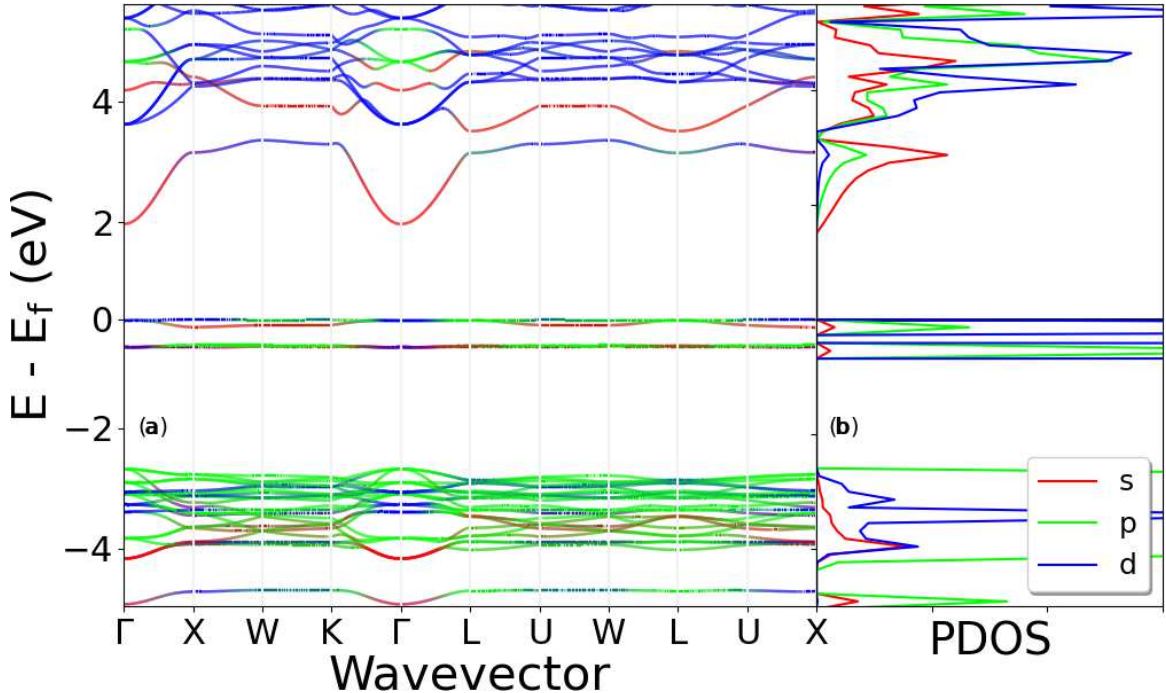}
\caption{(Color online.) Panel (a) shows the Kohn-Sham projected band structure (PBS), and panel (b) shows the PDOS for the doped K$_7$Cl$_8$:Cu system, Figure~\ref{doping}a.  s-, p-, and d-states are identified by the red, green , and blue lines, respectively.}
\label{bands5}
\end{figure}

The effect of the Cu atom on the electronic properties was analyzed. Figure~\ref{bands3}a shows the Kohn-Sham folded band structure for the doped K$_7$Cl$_8$:Cu symmetric system  shown in Figure~\ref{doping}a, and  along the same high-symmetry path used in the calculation for the pristine PC KCl. Figure ~\ref{bands3}b  shows a zoomed view of the Kohn-Sham folded band structure in the range of -4.0 to 0.5 eV, where the dispersionless character of the bands located at the (shifted) Fermi level can be clearly observed. The lowest calculated direct Kohn-Sham bandgap is 2.2 eV at $\Gamma$  the point.
The panels (a–h) of Figure~\ref{bands1}  shows the calculated Kohn-Sham folded band structure of SC K$_7$Cl$_8$:Cu as a function of the doping site and along with the same high-symmetry directions of the BZ used in the PC; for the octahedral square bipyramidal geometric-shape case shown in Figures~\ref{doping}a ; and for slightly distorted twisted octahedra square bipyramidal shape cases shown in Figure~\ref{octa}a (b$_1$-g$_1$), respectively. The lowest direct Kohn-Sham bandgap of the KCl:Cu obtained is 3.2 eV at the  point, for the anisotropic case of Figure~\ref{bands1}b as indicated by the red arrow and displayed in Table~\ref{geom2}. (For the cases Figures~\ref{bands1} (b-d) and (e-g), the Kohn-Sham bandgap is 3.2 eV. The anisotropy is indicative of the inhomogeneity in KCl:Cu material; and causes significant changes in the electronic properties of KCl:Cu system, opening the Kohn-Sham bandgap from 2.2 eV to 3.3 eV. In contrast, it has little impact on the shape of the band structure keeping the dispersion on the first conduction band. Figures~\ref{bands1} show the Kohn-Sham folded band structure for all eight cases to facilitate the comparison. To gain insight in the trend of the effect of Cu concentration in the electronic structure, we built the first and smallest cubic supercell that can be doped with Cu atom, with eight atoms basis (Cu, 3K, 4Cl), multiplicity 4, shown in Appendix~\ref{appendix:a}, Figure~\ref{cubic}b. The density of the Cu-doped KCl is 0.447 g/cm$^3$, considering the cubic SC with eight atoms. We show in Figure~\ref{bands4} the folded band structure calculated trough the ${\bf{k}}$-path corresponding a cubic cell. The Kohn-Sham bandgap is 0.80 eV at the $\Gamma$ point, and its character is indirect. Therefore, the energy bandgap decrease as the Cu concentration increases. We can infer if Cu concentration decreases, the band structure should increasingly resemble the pure KCl band structure.
The projected band structure (PBS) and PDOS of the doped Cu KCl system are depicted in Figure~\ref{bands5}. The PBS is color-coded in an informative manner: the color intensity corresponds to the degree of contribution of a particular orbital to the bands; thus visually, the contribution of orbitals to bands can be identified. The PBS of the KCl:Cu system, panel (a) Figure~\ref{bands5}  shows that the top-valence band, at the Fermi level and in the k-path line segments-W, K-,-L, and U-X, is formed by the contribution of the 3d states, and at the L k-points mainly by the 3p-states. Indeed, Figure~\ref{dos2}c indicates that the former contributions belong to the Cu atom and the latter to the chlorines atoms. The valence bands located at the energy range of -4 to -2 eV are mainly formed by Cl-3p states, as it is shown in Figure~\ref{bands5}. On the other hand, the 4s states of the Cu atom form the first conduction band at the K-$\Gamma$-L points. The color intensity of the first conduction band in the X-W-K and L-U-W-L-U-X segments indicates a mixture of s and p states (hybridization). The higher-energy conduction bands (blue lines) are composed mainly of the d electronic-states from the K atoms. Since critical optical transitions take place at the $\Gamma$  point and involve only the top valence band and the first conduction band, a 3d-4s optical transition of the Cu must occur, this is calculated in the optical properties section.
\section{Unfolded band structure of pristine K$_8$Cl$_8$ and Doped K$_7$Cl$_8$:Cu}
\begin{figure}[ht!]
  \begin{center}
  \setcounter{subfigure}{0} 
  \subfloat[][]{\includegraphics[scale=0.4]{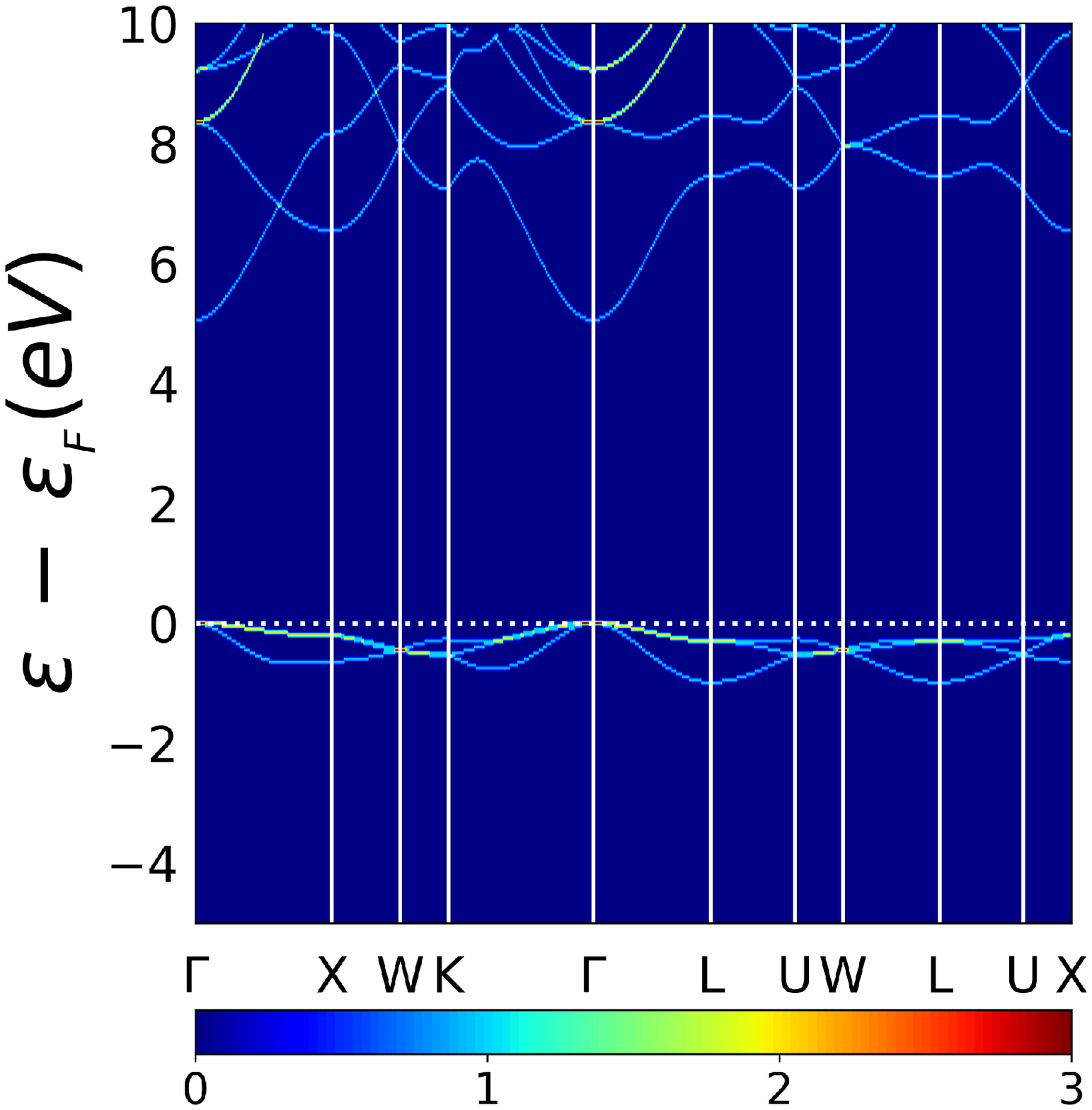}}\hspace*{\fill} 
  \subfloat[][]{\includegraphics[scale=0.4]{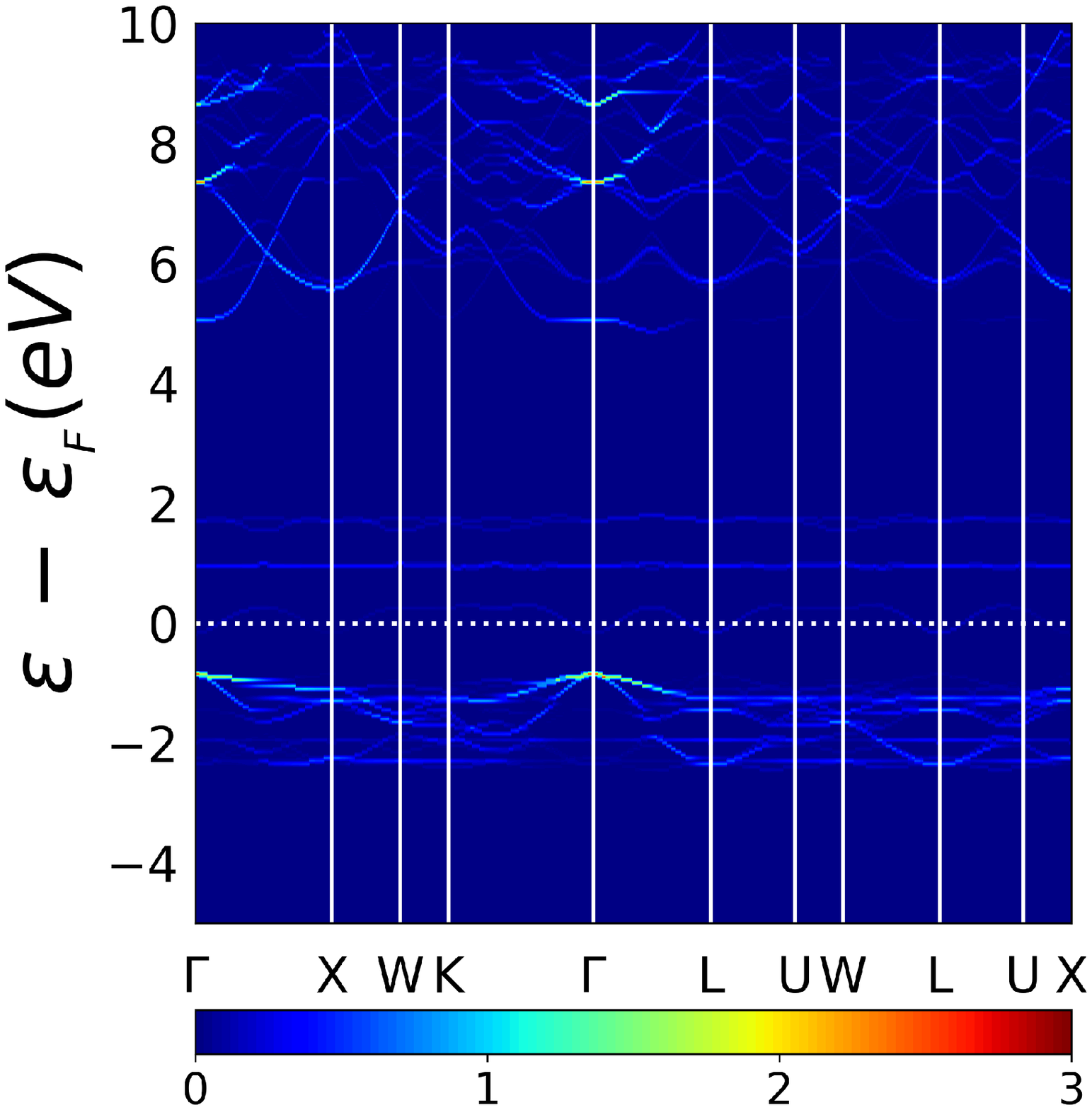}}\hspace*{\fill}
  \caption{(Color online.) Panel (a) shows the effective band structure (EBS) for pristine SC K$_8$Cl$_8$, Figure~\ref{cells}b and in Panel (b) shows the EBS for doped SC K$_7$Cl$_8$:Cu, Figure~\ref{doping}a, along with the same high-symmetry directions in the BZ and the same energy range. The color scale represents the number of PC bands crossing the energy interval (0.05 eV) at a given primitive wave vector.~\cite{medeiros,zhong,huang}}
  \label{unfold}
  \end{center}
\end{figure}

To gain a better understanding of the Cu atom doping effects on the KCl at the electronic level, we unfolded the band structure of the doped SC K$_7$Cl$_8$:Cu case  shown in Figures~\ref{doping}a. To perform a test and make a comparison with the doped system, we also unfolded the pristine SC K$_8$Cl$_8$. The panel (b) in Figure~\ref{unfold} shows the unfolded band structure for pristine SC K$_8$Cl$_8$ or the so-called effective band structure (EBS), introduced by Popescu~\cite{popescu2,popescu} and other researchers.~\cite{boykin,boykin2007,medeiros} The EBS can be directly compared to angle-resolved photoemission spectroscopy (ARPES) data.~\cite{warmuth} The EBS for the pristine SC K$_8$Cl$_8$ should give similar eigenvalues obtained from the band structure calculation employing PC KCl. We found the band structures of the PC and SC KCl to be identical due to the perfect translation of the primitive cells that built the pristine SC K$_8$Cl$_8$. To facilitate mutual comparison, the EBS shown in Figure~\ref{unfold} was computed within the same energy range and high-symmetry directions in the BZ employed in the band structure (blue lines) of PC KCl, as shown in Figure.~\ref{bands2}a The EBS of the doped SC K$_7$Cl$_8$:Cu is depicted in the  panel (a) in Figure.~\ref{bands1}b  As shown in this figure, the effect of Cu doping on the electronic structure of pure KCl is dramatic, the Cu doping greatly modifies the electronic structure of doped KCl, narrowing the energy bandgap and opening small gaps in the valence and conduction bands. The Fermi level has shifted to zero energy axes.
A detailed examination of Figure~\ref{unfold} indicates that Cu-doping of the KCl structure reduces the direct energy bandgap and that the unfolding procedure preserves the same direct character of the energy bandgap. Previous studies suggest that the folded bands fail to reproduce the indirect bandgap character.~\cite{deretzis,cartoixa} The cutoff of the energy bandgap is attributed to the appearance of empty bands with a dispersionless character located at 1.0 and 2.0 eV. The  panel (b) in Figure~\ref{unfold} shows the existence of a single valence band located at the Fermi level with small peaks and valleys; despite these, the band is quite flat, thus producing a pronounced peak in the PDOS. Further analysis of the valence bands reveals bunched-together bands in the energy range of −3.0 to −1.0 eV, showing many small energy gap openings that are not observed in the folded band structure of the SC K$_7$Cl$_8$:Cu, shown in Figure~\ref{bands1}. (black lines) Interestingly, similar energy gap openings were found in bi-graphene studies.~\cite{warmuth} Moreover, a small gap appeared in the bottom conduction bands at an energy of 5.0 eV, located at the $\Gamma$-L valley. A direct comparison between the bottom conduction band of pristine EBS (panel (a) in Figure~\ref{unfold} ) and the bottom conduction band of the doped EBS (panel (b) in Figure~\ref{unfold} ) shows that significant changes occurred in the k-$\Gamma$-L valley, which drastically changed the band dispersion and opened small gaps, as mentioned before.
\section{Optical properties}
\begin{figure}[ht!]
\centering
\includegraphics[scale=0.65]{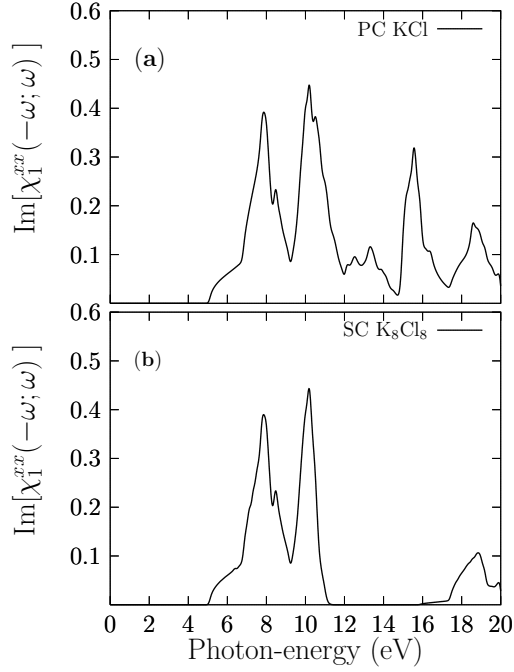}
\caption{(Color online.) The upper panel (a) shows the calculated $\mathfrak{Im}[\chi_1^{ab}(-\omega,\omega)]$  part corresponding to pristine PC KCl, with the FCC structure and two atoms in the base. The lower panel (b) shows the calculated $\mathfrak{Im}[\chi_1^{ab}(-\omega,\omega)]$ corresponding to pristine SC K$_8$Cl$_8$, (2 $\times$ 2 $\times$ 2 pristine PC KCl), with 16 atoms in the base.}
\label{optics1}
\end{figure}

Optical properties (OP) are a valuable source of atomic structural information, and their electronic band structure largely determines them.~\cite{onida,hybertsen} Once we obtained the optimized crystal structure of the KCl, the OPs were calculated by evaluating Equation~\ref{chi1}. Figure~\ref{optics1}a the $xx$ component of the imaginary  part of the electric susceptibility tensor for the pristine PC KCl in the energy range of 0 to 20 eV.
Figure~\ref{optics1}a shows the $\mathfrak{Im}[\chi_1^{ab}(-\omega,\omega)]$ part of the pristine PC. As shown in this figure, the  part starts to rise at 5.0 eV, and it has peaks located at 7.8, 10.2, 15.5, and 18.6 eV.
\begin{figure}[ht!]
\centering
\includegraphics[scale=0.65]{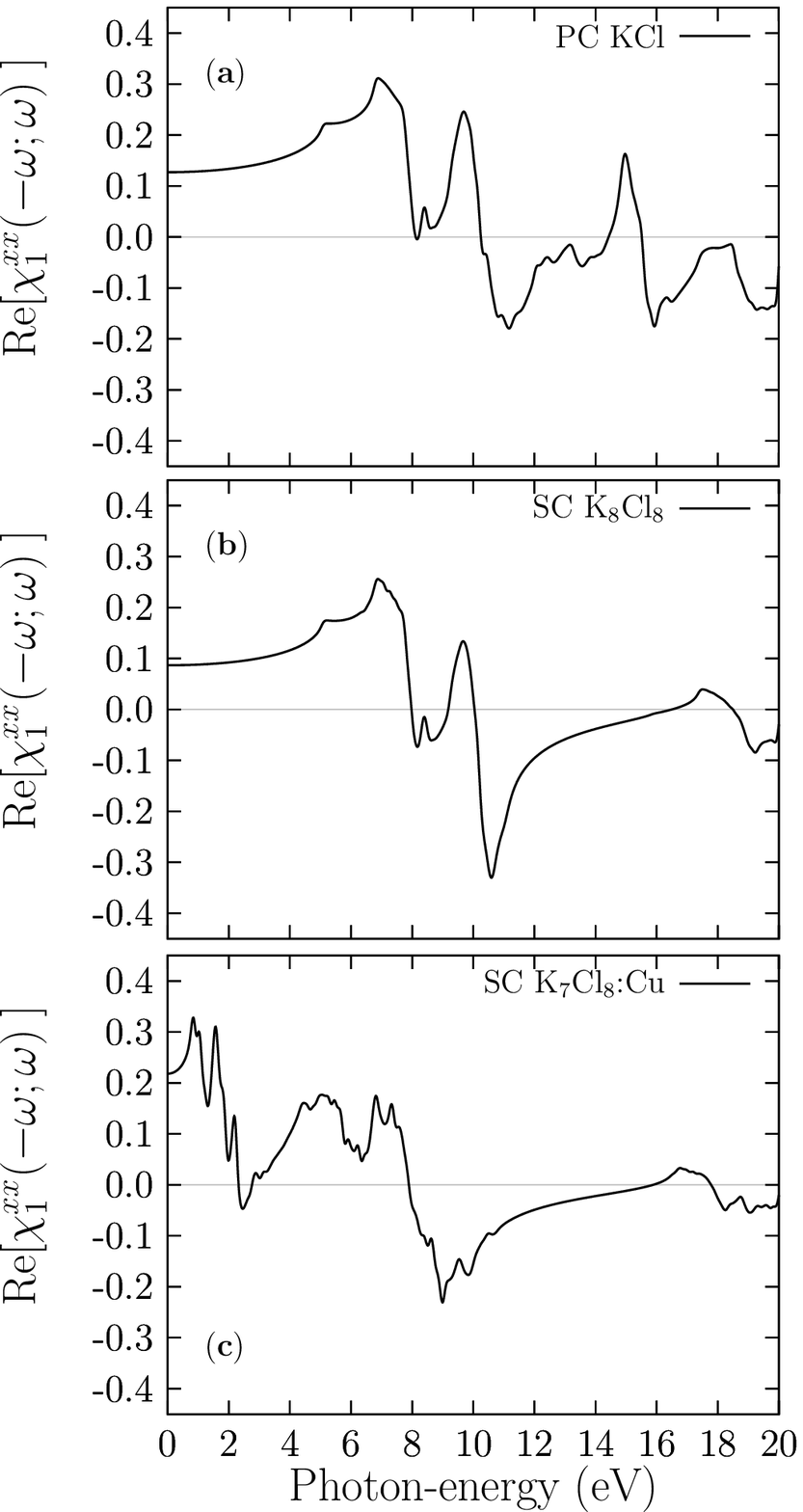}
\caption{(Color online.) Upper panel (a) shows the calculated real $\mathfrak{Re}[\chi_1^{ab}(-\omega,\omega)]$ part of the PC KCl. Middle panel (b) shows the real $\mathfrak{Re}[\chi_1^{ab}(-\omega,\omega)]$ part for the pristine SC K$_8$Cl$_8$ system. Lower panel (c) show the real $\mathfrak{Re}[\chi_1^{ab}(-\omega,\omega)]$  part of the doped SC K$_7$Cl$_8$ :Cu system, Figure~\ref{doping}a, and Figure~\ref{octa}a.}
\label{optics2}
\end{figure}
There are minor peaks located at 8.4 and 13.3 eV. At the onset of the signal, there is a shoulder in the energy range of 5.0 to 6.7 eV, which is mainly due to the direct transitions from the topmost valence bands (highest-energy valence bands) at the point. As the crystal structure of KCl is completely isotropic,(Figure~\ref{octa}a) the $xx$, $yy$, and $zz$ components are also isotropic. Hereafter, we only present the $xx$ component, and the scissor correction of 3.53 eV was applied at the end. Figure~\ref{optics1}b shows the $\mathfrak{Im}[\chi_1^{ab}(-\omega,\omega)]$ part or the pristine SC K$_8$Cl$_8$ that also start to raise at 5 eV and peaks located at 7.8 and 10.2 eV. The main difference between the PC and SC responses is located above 12.0 eV on the energy axis because the number of electronic bands necessary to converge the SC system is higher than that of the PC. Nonetheless, the signals are similar in the energy range of 5 to 10 eV as we expected.  Figure~\ref{optics2} shows the real  part of the pristine PC KCl,SC K$_8$Cl$_8$ and K$_7$Cl$_8$:Cu systems. The real $\mathfrak{Re}[\chi_1^{ab}(-\omega,\omega)]$  part provides information about the polarizability of the material. Figure~\ref{optics2}a  shows that the static value  is 0.1  for pristine PC KCl. From here, it starts to increase and reach a large peak with a value of 0.3 at 6.9 eV. After that, a rapid decrease is observed until it reaches zero. The negative values of the real  part are observed in the energy range of 10.28 to 14.43 eV and above 15.5 eV. At 10.28 eV, the real $\mathfrak{Re}[\chi_1^{ab}(-\omega,\omega)]$ part changes sign. The negative values of the real part are indicative of metallic behavior, whereas a change of sign in the real part indicates the occurrence of plasmonic resonances in the energy region where it crosses the energy axes with a positive slope. The real $\mathfrak{Re}[\chi_1^{ab}(-\omega,\omega)]$ part for the pristine SC shows in Figure~\ref{optics2}b has a similar behavior.
\section{Breakdown of the $\mathfrak{Im}[\chi_1^{ab}(-\omega,\omega)]$ part of pristine supercell (SC) KCl into different band contributions}
\begin{figure}[ht!]
\centering
\includegraphics[scale=0.65]{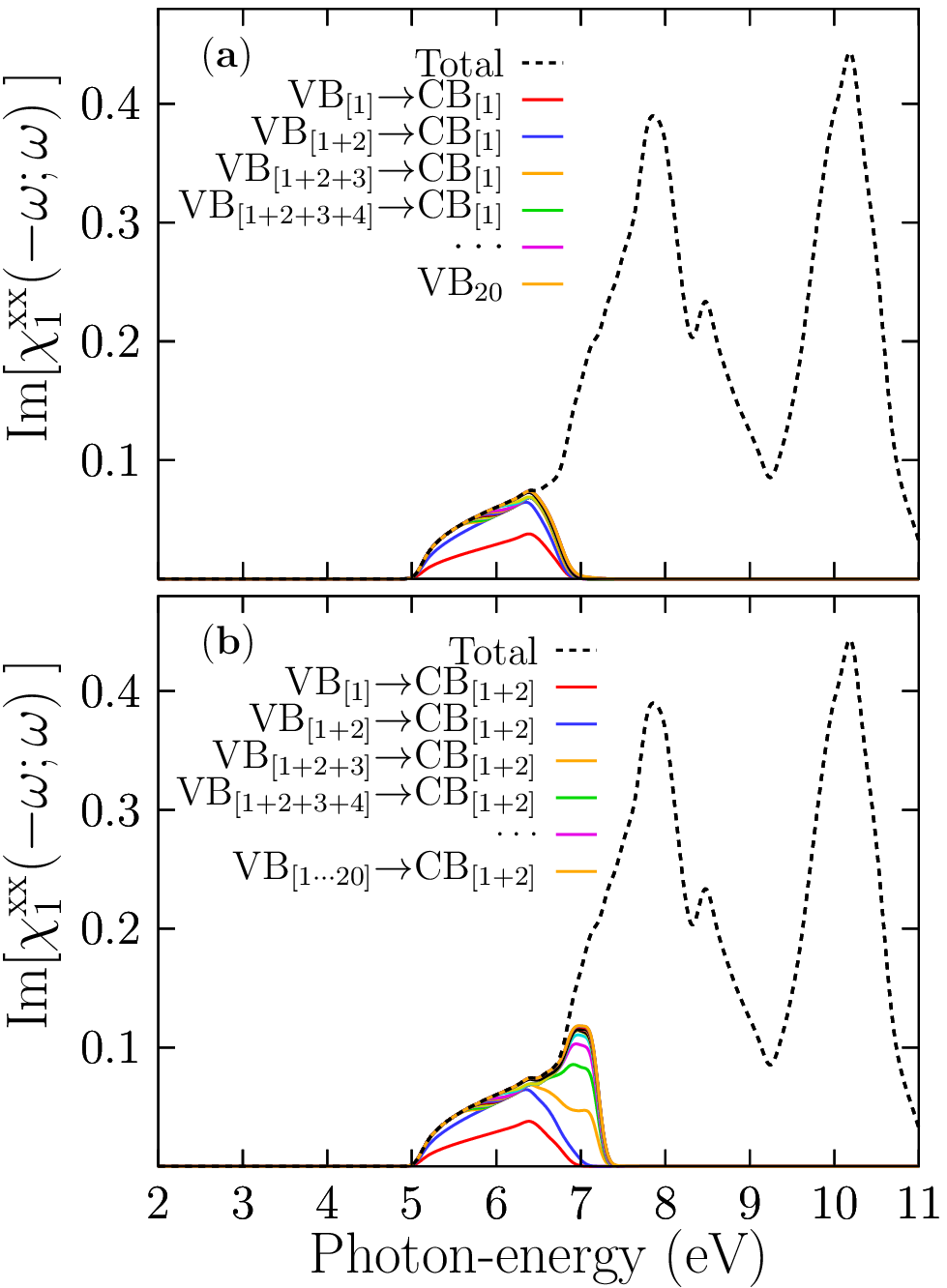}
\caption{(Color online.)Breakdown of the imaginary $\mathfrak{Im}[\chi_1^{ab}(-\omega,\omega)]$ part of the SC K$_8$Cl$_8$ into different band contributions. The upper panel (a) shows the decomposition of the part by only considering the 20 highest valence band (VB) transitions to only the first conduction band (CB), (\emph{e.g.} VB$_{[1]}$ $\rightarrow$ CB$_{[1]}$,VB$_{[1+2]}$$\rightarrow$CB$_{[1]}$,VB$_{[1+2+3]}$$\rightarrow$CB$_{[1]}$). The lower panel (b) shows the decomposition of the $\mathfrak{Im}[\chi_1^{ab}(-\omega,\omega)]$ part only considering the 20 highest VB transitions to only the first and second CBs, (\emph{e.g.} VB$_{[1]}$$\rightarrow$CB$_{[1+2]}$, VB$_{[1+2]}$$\rightarrow$CB$_{[1+2]}$,   VB$_{[1+2+3]}$$\rightarrow$CB$_{[1+2]}$). The red solid line is the VB $_{[1]}$ $\rightarrow$ CB$_{[1]}$ transition. The suspension points, pink solid line, stand for terms not listed, but which follow the sequence. The black dotted line in both panels represents the  part, taking into account all VBs and CBs considered in the calculation.}
\label{optics3}
\end{figure}

Figure 10a shows the breakdown of the $\mathfrak{Im}[\chi_1^{ab}(-\omega,\omega)]$ part, taking into account the transitions within an iterated cumulative sum (ICS) of all the valence bands, assigning to the top valence band the index 1 (VB$_1$), the sub-top valence band 2 (VB$_2$), and so on (e.g., VB$_{[1]}$, VB$_{[1+2]}$) against the single bottom conduction band (CB$_1$). In this in-house developed scheme, the first ICS corresponds to the contributions to the  part due to the top valence and bottom conduction bands. The red line in Figure 10a clearly shows that the contribution to the onset of the signal up to 6.5 eV is 50{\%}. The second ISC, which takes into account the top valence and sub-top valence bands against the single bottom conduction band, is depicted by the solid blue line in Figure 10a, which clearly shows that this ICS contributes to the  part by almost 95{\%} for the onset of the signal. The rest of the ICS contributes only about 5{\%}. Twenty of the ICSs are plotted and shown in Figure~\ref{optics3}. Our calculated band contribution breakdown indicates that we only need the two topmost valence bands and the bottom conduction band to build up the onset of the signal of up to 6.5 eV with an accuracy of 95{\%}. Moreover, based on the PDOS, the Cl-3p states mainly form the two topmost valence bands, and the bottom conduction band is composed of a mixture of Cl-s and K-s states. Therefore, the onset of the signal up to 6.5 eV is due to the direct transitions involving Cl-3p, Cl-4s, and K-4s states at the $\Gamma$ point. 
Figure~\ref{optics3}b shows the breakdown of the $\mathfrak{Im}[\chi_1^{ab}(-\omega,\omega)]$ part with an ICS considering all valence bands against the two bottom-most conduction bands. The figure shows that the contribution of the first ICS, depicted by the solid blue line, considers the transitions from the top valence band to the two bottom-most conduction bands, which are similar to those when we consider a single bottom conduction band. The third ICS, depicted as a solid yellow line in Figure~\ref{optics3}b, builds the large peak located at 7.8 eV. In summary, and to solve the first large peak located at 7.8 eV and peaks located at higher energies, we need to take higher conduction bands into account.
\begin{figure}[ht!]
\centering
\includegraphics[scale=0.65]{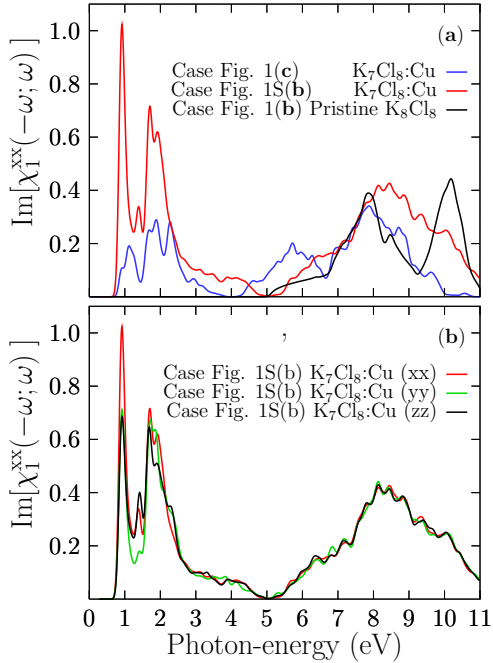}
\caption{(Color online.)In the upper panel (a) the solid blue line represents the $\mathfrak{Im}[\chi_1^{ab}(-\omega,\omega)]$ part of the SC K$_7$Cl$_8$:Cu of the isotropic case shows in Figure~\ref{doping}a. The solid red line represent the $\mathfrak{Im}[\chi_1^{ab}(-\omega,\omega)]$ part of the SC K$_7$Cl$_8$:Cu of the anisotropic case is shown in Figure~\ref{doping}b. The solid black line represents the $\mathfrak{Im}[\chi_1^{ab}(-\omega,\omega)]$ part of the pristine SC K$_8$Cl$_8$ system. The lower panel (b) shows the $\mathfrak{Im}[\chi_1^{xx}(-\omega,\omega)]$,$\mathfrak{Im}[\chi_1^{yy}(-\omega,\omega)]$, and $\mathfrak{Im}[\chi_1^{zz}(-\omega,\omega)]$ parts of the SC K$_7$Cl$_8$:Cu, the octahedral square bipyramidal shape case is shown in Figure~\ref{doping}b. Notice the optical anisotropy is induced by the Cu doping position. The energy range is 0 to 11 eV.}
\label{optics4}
\end{figure}
\section{The $\mathfrak{Im}[\chi_1^{ab}(-\omega,\omega)]$ part of the doped K$_7$Cl$_8$:Cu}
The solid blue line shown in Figure~\ref{optics4}a depicts the $\mathfrak{Im}[\chi_1^{ab}(-\omega,\omega)]$ part that corresponds to the doped SC K$_7$Cl$_8$ octahedral square pyramidal geometric-shape shown in Figure~\ref{octa}a, the solid black line corresponds to the  $\mathfrak{Im}[\chi_1^{ab}(-\omega,\omega)]$ part of the pristine SC K$_8$Cl$_8$ system. The solid red line corresponds to the doped SC K$_7$Cl$_8$:Cu slightly distorted-twisted octahedra square bipyramidal shape shown in Figure.~\ref{octa}b We have included the three signals on the same graphic to compare them. We first focus on the isotropic case, solid blue line. Figure~\ref{optics4}a  shows that doping KCl with Cu had a drastic effect on in its optical properties. The main peak starts to rise at 0.9 eV and ends at 4.0 eV and is composed of four sub-peaks located at 1.1, 1.87, 2.3, and 2.9 eV. In the energy range of 4.0 to 6.7 eV, there is a large peak located at 5.7 eV with little ripples. Above the energy range of 6.7 eV, the signals and peaks for the doped and pristine systems are similar, and those peaks are located at an energy of 7.8 eV. The larger peak observed in the pristine signal around 10 eV, and not observed in the doped system, is caused by the number of conduction bands considered in the calculation. Figure~\ref{optics2}c shows the $\mathfrak{Re}[\chi_1^{ab}(-\omega,\omega)]$ part for the doped SC K$_7$Cl$_8$:Cu. It also clearly shows that the static value of the  part is 0.22, , which is slightly more than double for the pristine case of 0.1. Figure~\ref{optics2}c  also shows that the largest peak, with a value of 0.3, is located at 0.8 eV, followed by a rapid decrease until 0.15 at 1.31 eV, after starting to rise to form a large peak located at 1.6 eV. From here, it begins to decrease until it crosses zero at 2.3 eV. At this energy value, the real part changes sign, indicating an occurrence of plasmonic resonances. Negative values of the part are observed in the energy range of 2.30 to 2.70, 7.90 to 15.9 eV, and above 17.8 eV. Comparing the  part of the pristine and the  part of the doped case, we find significant differences, similar to those found in the imaginary parts. As a consequence, the refraction index of the doped SC K$_7$Cl$_8$:Cu, which is related to the static value of the dielectric function, can increase at around 70{\%} compared with the pristine case. Figure 11 lower panel (b)~\ref{optics4}  shows the  part $\mathfrak{Im}[\chi_1^{ab}(-\omega,\omega)]$ that corresponds to the doped SC K$_7$Cl$_8$:Cu slightly distorted-twisted octahedral square bipyramidal shape shown in Figure~\ref{octa}b. The optical anisotropy is indicative of the inhomogeneous KCl:Cu, which results from the asymmetrically Cu-Cl bonds in the octahedra of the unit cell. Panel (b) of  Figure panel~\ref{optics4}  shows the imaginary $\mathfrak{Im}[\chi_1^{ab}(-\omega,\omega)]$  part of the diagonal elements of the dielectric tensor. The solid red line, solid green line, and solid black line are the $xx$, $yy$, and $zz$ elements. They reveal a considerable anisotropy in the energy range up to 3 eV. Beyond this point, the imaginary parts of the diagonal components of the dielectric tensor $xx$, $yy$, and $zz$ are equal to each other; this is because the transitions contributing to this part of the spectrum come from the Cl and K atoms. At the onset of the signal (solid red line), the curve rapidly increases because the number of transitions contributing to the signal’s onset comes from the Cu atoms. Thus, in the x-axis direction, there is a larger concentration of Cu atoms than in the y-axis and z-axis directions. The arrangement of six chlorine anions around the centered copper cation set a charge of –1 at each vertex, which causes the d-Cu states to split into two groups with different energies. The energy difference between the two groups of orbitals is known as the crystal field splitting energy. On the other hand and from the electronic point of view, the magnitude of the splitting depends on the charge of the Cu atom, and a small change in the crystal field splitting energy is related to a distortion of the octahedron. The distorted-twisted octahedral square bipyramidal shape induces a difference in the crystal field splitting energy compared to that of the perfect octahedral square bipyramidal shape. Therefore, the origin of the anisotropy comes from the difference in the crystal field-splitting energy between the distorted-twisted and the perfect octahedral square bipyramidal geometric-shapes. Moreover, the distorted system -Jahn-Teller effect- could be more stable than the undistorted one. In summary, the doping Cu atoms position has an enormous impact on the crystal field-splitting energy, and as a consequence, the optics and electronic properties change significantly.
\section{Breakdown of the $\mathfrak{Im}[\chi_1^{ab}(-\omega,\omega)]$ part of the doped SC K$_7$Cl$_8$:Cu into band contributions}
\begin{figure}[ht!]
\centering
\includegraphics[scale=0.65]{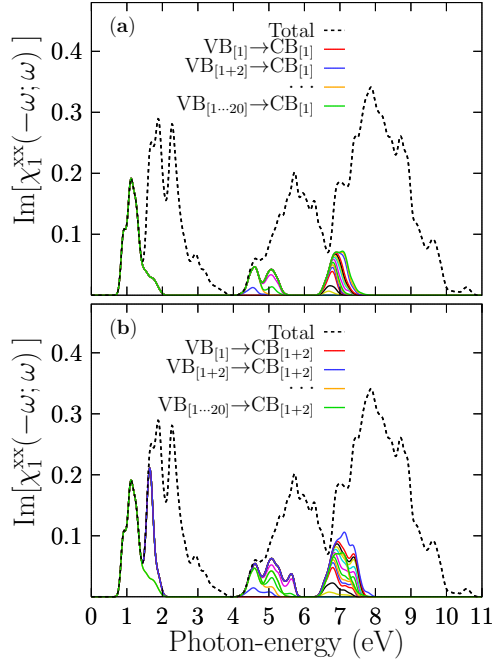}
\caption{(Color online.) Breakdown of the $\mathfrak{Im}[\chi_1^{ab}(-\omega,\omega)]$  part of the K$_7$Cl$_8$:Cu, Figure.\ref{doping}a, into different band contributions. The upper panel (a) shows the $\mathfrak{Im}[\chi_1^{ab}(-\omega,\omega)]$  part, taking into account all valence bands with a single bottom conduction band. The lower panel (b) shows the $\mathfrak{Im}[\chi_1^{ab}(-\omega,\omega)]$  part, taking into account all valence bands with the two bottom-most conduction bands. The suspension points, $\cdot  \cdot \cdot$  yellow solid line, stand for terms not listed, but which follow the sequence. The black dotted line represents the total $\mathfrak{Im}[\chi_1^{ab}(-\omega,\omega)]$ part, considering all valence and conduction bands available.}
\label{optics5}
\end{figure}

To elucidate the transitions contributing to the peaks of the $\mathfrak{Im}[\chi_1^{ab}(-\omega,\omega)]$ part  that are located in the energy range of interest, we depict, in Figure~\ref{optics5}, the computed breakdown of the $\mathfrak{Im}[\chi_1^{ab}(-\omega,\omega)]$ part into band contributions (see also the video in the supplementary material).
In Figure~\ref{optics4}a , the first ICS that involves the single top valence band (VB$_{[1]}$) and single bottom conduction band (CB$_{[1]}$) accounts for the entire peak of the $\mathfrak{Im}[\chi_1^{ab}(-\omega,\omega)]$ part located at 1.1 eV. Following the calculated PDOS presented in Figure~\ref{dos2}c, the top valence band is dispersionless and is composed of Cu-3d states, while the bottom conduction band is mainly composed of Cu-s states. Thus, the first peak of the $\mathfrak{Im}[\chi_1^{ab}(-\omega,\omega)]$ part , located at 1.1 eV, is caused by the transitions of Cu-3d to Cu-4s, mainly at the point. For Cu-s states at around 2.2 eV, this Figure also shows a PDOS with a low density of states, implying a broader bottom conduction band. In contrast, the same PDOS shows a very high density of states located at the Fermi level, indicating a relatively narrow valence band. Our calculations show that the intraband transitions take place from a dispersionless band composed of Cu-3d states to the bottom conduction band formed by Cu-4s states. Our calculated Kohn-Sham bandgap for the pristine case is 5.07 eV, which is an underestimation of 41{\%} compared with the experimental value of 8.6 eV. If we consider this 59{\%} to estimate the correct bandgap of the doped system, we obtain 3.90 eV; therefore, the first peak of the should be located at 3.17 eV, in excellent agreement with previous works.~\cite{myasnikova2} Experimentally, it has been corroborated that there is a transition for copper, located at 2.2 eV, which explains why the copper has a reddish color.~\cite{fox} Considering our calculations and the shift to 3.17 eV of the first absorption peak, the doped system should have a violet color due to the transitions between the top valence band, formed by Cu-3d states, and the bottom conduction band, formed by Cu-4s states. The second ICS that involves the topmost and second-topmost valence bands (VB$_{[1+2]}$) and the single bottom conduction band (CB$_{[1]}$) is depicted in the solid blue line of Figure~\ref{optics4}a. This transition contributes to the first peak located at 1.1 eV and the peak at 4.3 eV. Therefore, the peak located at 5.7 eV is also due to the Cu-3d states. The ICS involving valence bands with 3 to 5 indices account for the small peaks located at 5.7 eV, and those that involve valence bands with 7 to 40 indices against the single bottom conduction band contribute to the most significant peak, located at 7.8 eV. According to the calculated PDOS shown in Figure~\ref{dos2}c, these valence bands are located in the energy range of −2.0 to −4.0 eV and are mainly composed of Cl-3p states. This peak begins with transitions from Cl-3p states to the bottom conduction band, which is composed of a mixture of Cu-4s.
To gain more insight into the optical absorption details, we now break down the part considering the two bottom-most conduction bands against all the valence bands. The first ICS transition, taking into account the single top valence and two bottom conduction bands, is depicted as a solid red line in Figure~\ref{optics4}b It provides the sole contribution to the first peak of the part located at 1.1 eV and contributes in part to a peak located at 1.87 eV. The behavior is similar to the case when we take a single bottom conduction band. Indeed, the four peaks of the  part located at 1.1, 1.87, 2.3, and 2.9 eV involve transitions from the two topmost valence bands to the bottom conduction bands. The PDOS indicates that those transitions involve Cu-3d and K-3d states. In summary, the first peak located at 1.1 eV is initiated by the transitions between the top valence bands, formed by pure Cu-3d states, to the bottom conduction band, formed by Cu-4s states. The second peak, located at 1.87 eV, is associated with transitions from the topmost valence bands to the second bottom bands. Therefore, it is contributed to by transitions from the top valence band formed by pure Cu-3d states to the second conduction band formed by K-3d states. The onset of the peak located at 5.7 eV is initiated by the transitions between bands formed mainly of Cl-3p states to conduction bands formed of Cu-4s. Indeed, transitions to conduction bands formed by K-3d states (located at 4 eV) must also occur to form this peak.
\section{Conclusions}
In summary, we employed the band structure unfolding technique to study the electronic properties and break down the imaginary $\mathfrak{Im}[\chi_1^{ab}(-\omega,\omega)]$ part of the dielectric susceptibility of the pristine KCl and doped KCl:Cu systems to gain fundamental insights into the effect of copper doping on the electronic structure and optical properties of pure KCl. Additionally, we considered different doping concentration and doping positions, where the Cu dopant occupies all the substitutional sites replacing host K cations. Substitutional doping leads to the distortion of the atomic structure near impurity atoms. The arrangement of six chlorine anions around the centered copper cation sets a charge of -1 at each vertex, which causes the d-Cu states to split into two groups with different energies. The distorted-twisted octahedral square bipyramidal geometric-shape induces a difference in the crystal field-splitting energy compared to that of the perfect octahedral square bipyramidal geometric-shape. Therefore, the origin of the anisotropy comes from the difference in the crystal field-splitting energy between the distorted-twisted and the perfect octahedral square bipyramidal geometric-shapes. Moreover, the distorted system -Jahn-Teller effect- could be more stable than the undistorted one. The doping Cu atom’s position has an enormous impact on the crystal field splitting and as a consequence, the optics and electronic properties change significantly. To study the changes in the band structure, we performed a direct comparison with the band structure of the pure KCl. In addition, we unfolded the band structure of the pure SC KCl, exactly recovering the band structure of the PC KCl and analyzed the projected band structure of the doped KCl:Cu Our findings show significant differences between the unfolded band structure of the Cu-doped KCl and the unfolded band structure of pure KCl. (also PC-band structure) For the doped system, the bandgap was drastically reduced. Moreover, a dispersionless electronic band associated to Cu-3d states appears at the Fermi level, and the conduction bands above the Fermi level are attributed to the Cu-4s electronic states. There is also less dispersion in the bottom conduction band of the doped system and a large dispersion in the bottom conduction band for pristine KCl. In general, the main differences between the folded and the unfolded band structures of the doped KCl are the drastic change in the overall dispersion and the opening of small energy gaps in the bands, which were revealed by the unfolding procedure. The unfolding method was proven useful in the study of doped materials or those containing vacancies. The projected band structure shows that there are Cu-3d states at the $\Gamma$ point that belongs to the top valence band; also there are Cu-4s states in the $\Gamma$ point, that belong to the first conduction band; as a consequence all the transitions that take place at the point and between top valence and first conduction bands are transitions 3d-4s of the Cu atom. To gain insight in the effect of Cu concentration in the electronic structure, we built the first and smallest cubic supercell that can be doped with a Cu atom and calculated the folded band structure. We found that the Kohn-Sham bandgap is 0.80 eV and its character is indirect. The trend is that the energy bandgap decrease as the Cu atom concentration increases and vice versa. To study the optical properties, we broke down the  part into band contributions. The sum-over-states formalism allows one to decompose band-to-band contribution through partial band-index summation, the bands of which significantly contribute to . This methodology, proposed here together with PDOS, allows us to identify and analyze the origin of the peaks presented in the $\mathfrak{Im}[\chi_1^{ab}(-\omega,\omega)]$ part. Our findings for the doped case show that the first peak of $\mathfrak{Im}[\chi_1^{ab}(-\omega,\omega)]$ located at 1.1 eV is due to transitions between the top valence band, which consists solely of Cu-3d states, to the bottom conduction band, formed of Cu-4s states. Therefore, the first peak is attributed to the transitions from Cu-3d to Cu-4s states. The second peak, located at 1.87 eV, is due to transitions between the top valence band and bottom conduction band. In addition, transitions between the top valence band and the sub-bottom conduction band were ascribed to K-3d states, i.e., the second peak of  located at 1.87 eV is due to an admixture of electronic transitions from Cu-3d to Cu-4s states and from Cu-3d to K-3d states. We explain the origin of the onset of the signals and demonstrate that the methodology of decomposing the imaginary $\mathfrak{Im}[\chi_1^{ab}(-\omega,\omega)]$ part into band contributions is a powerful tool in achieving a better understanding of the transitions participating in the optical response peaks. We demonstrated that The doping Cu atom’s position has an enormous impact on the \emph{crystal field splitting energy} and as a consequence, the optics and electronic properties change significantly.
\section{Acknowledgments}
Computational resources supporting this work were provided by a)\emph{ACARUS}, Dra. Carmen Heras, and L.C.C. Daniel Mendoza, through the High-Performance Computing Area of the University of Sonora. b) \emph{ELBAKYAN} and \emph{PAKAL} supercomputers.  
\section{Conflicts of Interest} The authors declare no conflict of interest.
\bibliography{bibliografia}
\appendix
\section{Pure KCl SC cell on the basis of eight atoms}
\label{appendix:a}
\begin{figure}[ht]
  \begin{center}
  \setcounter{subfigure}{0} 
  \subfloat[][]{\includegraphics[scale=0.35]{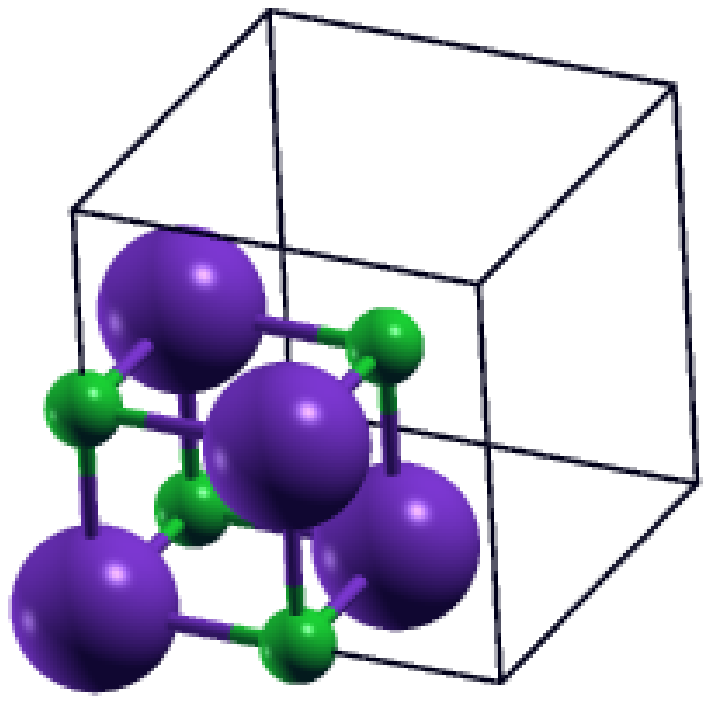}}\hspace*{\fill} 
  \subfloat[][]{\includegraphics[scale=0.35]{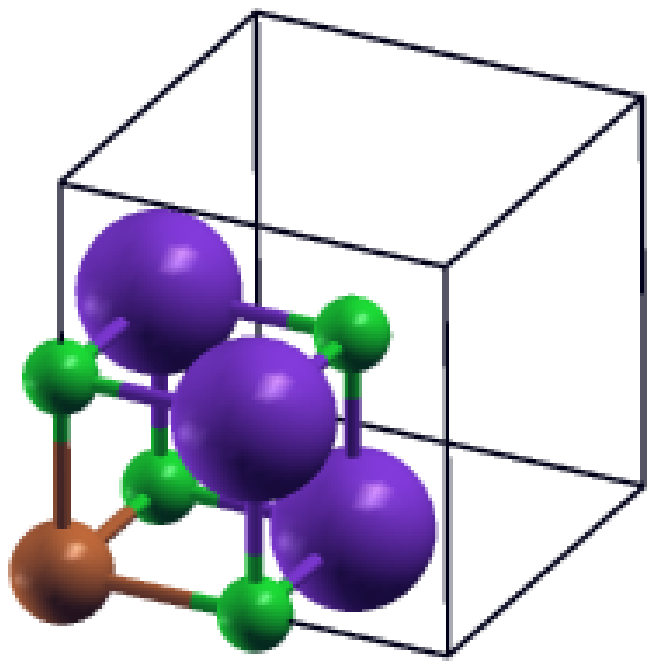}}\hspace*{\fill}
  \caption{(Color online.) Cubic supercell of pure KCl with eight atoms at the base, (a) clean and (b) doped.  The relation between cells FCC-PC, shows in Figure 1a, and cubic SC with eight atoms basis, is a geometric factor of 1.4142. The computed lattice constant in the FCC-PC  4.507~\AA,~we can calculate the lattice constant for cubic-SC employing the factor  1.4142.  The violet- and green-colored spheres represent the potassium and chlorine atoms, respectively.}
  \label{cubic}
  \end{center}
\end{figure}
\section{List of non-equivalent atoms in the unit cell.}
\label{appendix:b}
\begin{table}[ht]
  \caption{List of non-equivalent atoms in the unit cell (cryst. Coords.) of the  KCl:Cu system,~Figure\ref{doping}a}
  \label{list}
\centering
\begin{tabular}{lccc}
  \toprule
  \hline
  \hline
\textbf{Atom}	& \textbf{x}	& \textbf{y} & \textbf{x}  \\
\hline
 Cu  &  0.0000000000	& 0.0000000000  & 0.0000000000\\
 K   &  0.5000000000	& 0.0000000000  & 0.0000000000\\
 K   &  0.5000000000 	& 0.5000000000  & 0.5000000000\\
 Cl  &  0.2500000000	& 0.2500000000  & 0.2500000000\\
 Cl  &  0.7765725826	& 0.2234274174  & 0.2234274174\\
\hline
\end{tabular}
\end{table}
\begin{table}[ht]
  \caption{List of atoms in the unit cell (cryst. coords.)  of the KCl:Cu system,~Figure\ref{doping}a. Molecular formula: Cu(1) K(6) K(1) Cl(2) Cl(6). Number of non-equivalent atoms in the unit cell: 5. Number of atoms in the unit cell: 16. Number of electrons: 298.}
  \label{list2}
\centering
\begin{tabular}{lccc}
  \toprule
  \hline
  \hline
  \textbf{Atom}	& \textbf{x}	& \textbf{y} & \textbf{x}  \\
\hline
 Cu &0.0000000000  & 0.0000000000 &0.0000000000 \\
 K  &0.5000000000  & 0.0000000000 &0.0000000000 \\
 K  &0.0000000000  & 0.0000000000 &0.5000000000\\
 K  &0.0000000000  & 0.5000000000 & 0.5000000000\\
 K  &0.5000000000  & 0.5000000000 & 0.0000000000\\
 K  &0.5000000000  & 0.0000000000 & 0.5000000000\\
 K  &0.0000000000  & 0.5000000000 & 0.0000000000\\
 K  &0.5000000000  & 0.5000000000 & 0.5000000000\\
 Cl &0.2500000000  & 0.2500000000 & 0.2500000000\\
 Cl &0.7500000000  & 0.7500000000 & 0.7500000000\\
 Cl &0.7765725826  & 0.2234274174 & 0.2234274174\\
 Cl &0.2234274174  & 0.7765725826 & 0.7765725826\\
 Cl &0.7765725826  & 0.7765725826 & 0.2234274174\\
 Cl &0.2234274174  & 0.2234274174 & 0.7765725826 \\
 Cl &0.2234274174  & 0.7765725826 & 0.2234274174\\
 Cl &0.7765725826  & 0.2234274174 & 0.7765725826\\
\hline
\end{tabular}
\end{table}
\begin{table}[ht]
  \caption{Lattice vectors (bohr) of the KCl:Cu system,~Figure\ref{doping}a. The lattice parameters (bohr): 15.855582  15.855582  15.855582, The lattice parameters (ang): 8.390413  8.390413  8.390413, and The lattice angles (degrees): 60.000  60.000  60.000..}
  \label{list3}
\centering
\begin{tabular}{lccc}
  \toprule
  \hline
  \hline
  \textbf{Atom}	& \textbf{x}	& \textbf{y} & \textbf{x}  \\
\hline
 a  & 15.8555818239 & 0.0000000000   & 0.0000000000 \\
 b  & 7.9277909119  & 13.7313366513  & 0.0000000000 \\
 c  & 7.9277909119  & 4.5771122171   & 12.9460283478 \\
 \hline
\end{tabular}
\end{table}
\section{The PC  band structure with applied scissors correction.}
\label{appendix:c}
\begin{figure}[ht]
  \begin{center}
  \includegraphics[scale=0.65]{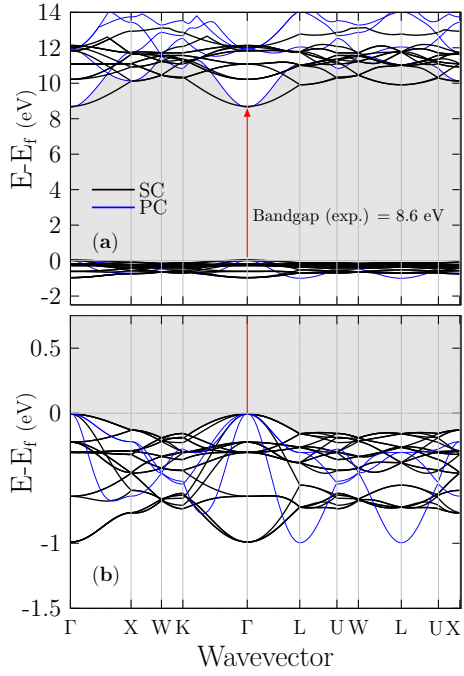}
  \caption{(Color online.)we show the PC band structure with applied scissors correction of 3.53 eV.}
  \label{scissors}
  \end{center}
\end{figure}
\end{document}